\documentclass[useAMS,usenatbib]{mn2e}

\usepackage{psfig,epsfig}
\usepackage{mathrsfs}
\usepackage{amssymb}
\usepackage{epsfig}
\usepackage{marvosym}
\usepackage{subfigure}
\usepackage{fancyhdr}
\usepackage{rotating}
\usepackage{color}

\def\degr{\hbox{$^\circ$}}

\def\gsim{\mathrel{\hbox{\rlap{\lower.55ex \hbox {$\sim$}}
                   \kern-.3em \raise.4ex \hbox{$>$}}}}
\def\lsim{\mathrel{\hbox{\rlap{\lower.55ex \hbox {$\sim$}}
                   \kern-.3em \raise.4ex \hbox{$<$}}}}

\def\he{\hbox{He\,{\sc i} $\lambda$5875}}

\def\EGAPS{\hbox{\sl EGAPS }}
\def\UVEX{\hbox{\sl UVEX }}
\def\IPHAS{\hbox{\sl IPHAS }}

\def\SDSS{\hbox{\sl SDSS }}

\def\RL1{\hbox{{$R_{L_{1}}$}}}

\title[A determination of the space density and birth rate of hydrogen-line (DA) white dwarfs in the Galactic Plane]
{A determination of the space density and birth rate of hydrogen-line (DA) white dwarfs in the Galactic Plane, based on the UVEX survey}
\author[Kars Verbeek et al.]{{Kars Verbeek$^{1}$\thanks{E-mail:k.verbeek@astro.ru.nl}, 
Paul J. Groot$^{1}$,
Gijs Nelemans$^{1}$,
Simone Scaringi$^{1,3}$,}
\newauthor{
Ralf Napiwotzki$^{2}$,
Janet E. Drew$^{2}$,
Danny Steeghs$^{4}$,
Jorge Casares$^{5}$,}
\newauthor{
Jesus M. Corral-Santana$^{5,6}$
Boris T. G{\"a}nsicke$^{4}$,
Eduardo Gonz{\'a}lez-Solares$^{7}$,}
\newauthor{
Robert Greimel$^{8}$,
Ulrich Heber$^{9}$,
Mike J. Irwin$^{9}$,
Christian Knigge$^{10}$,}
\newauthor{
Nicholas J. Wright$^{2}$
and Albert A. Zijlstra$^{11}$}\\
$^{1}$Department of Astrophysics/IMAPP, Radboud University Nijmegen,
  P.O. Box 9010, 6500 GL Nijmegen, The Netherlands\\
$^{2}$Centre for Astronomy Research, Science \& Technology Research
  Institute, University of Hertfordshire, Hatfield, AL10 9AB, UK\\
$^{3}$Instituut voor Sterrenkunde, KU Leuven, Celestijnenlaan 200D, B-3001 
  Leuven, Belgium\\
$^{4}$Physics Department, University of Warwick, Coventry, CV4 7AL,
  UK\\
$^{5}$Instituto de Astrof\'{\i}sica de Canarias, Via Lactea, s/n
  E-38205 La Laguna (Tenerife), Spain\\
$^{6}$Departamento de Astrof\'{\i}sica, Universidad de La Laguna, 
  La Laguna E-38205, S/C de Tenerife, Spain\\
$^{7}$Cambridge Astronomy Survey Unit, Institute of Astronomy, University of
  Cambridge, Madingley Road, Cambridge, CB3 0HA, UK\\
$^{8}$Institut f\"ur Physik, Karl-Franzen Universit\"at Graz,
Universit\"atsplatz 5, 8010 Graz, Austria\\
$^{9}$Dr. Remeis-Sternwarte Bamberg, Universit\"at Erlangen-N\"urnberg,
  Sternwartstrasse 7, D-96049 Bamberg, Germany\\
$^{10}$School of Physics and Astronomy, University of Southampton,
  Southampton, Hampshire, SO17 1BJ, UK\\
$^{11}$ Jodrell Bank Centre for Astrophysics, Alan Turing Building, 
  University of Manchester, M13 9PL, UK
}

\begin{document}

\date{Accepted for publication in MNRAS}

\pagerange{\pageref{firstpage}--\pageref{lastpage}} \pubyear{2013}

\maketitle

\label{firstpage}

\begin{abstract}
We present a determination of the average space density and birth rate of hydrogen-line 
(DA) white dwarfs within a radius of 1 kpc around the Sun, based
on an observational sample of 360 candidate white dwarfs 
with $g$$<$19.5 and $(g-r)$$<$0.4, selected from the 
UV-excess Survey of the Northern Galactic Plane (\UVEX), 
in combination with a theoretical white dwarf population
that has been constructed to simulate the observations, including the
effects of reddening and observational selection effects.
The main uncertainty in the derivation of the white dwarf space
density and current birth rate lies in the absolute photometric
calibration and the photometric scatter of the observational data, 
which influences the classification method on colours,
the completeness and the pollution. 
Corrections for these effects are applied.
We derive an average space density of hydrogen-line (DA) white dwarfs with
$T_{\rm eff}$$>$10\,000K ($M_{V}$$<$12.2) of
(3.8 $\pm$ 1.1) $\times$ 10$^{-4}$ pc$^{-3}$, 
and an average DA white dwarf birth rate over the last 7$\times$10$^7$ years of 
(5.4 $\pm$ 1.5) $\times$ 10$^{-13}$ pc$^{-3}$yr$^{-1}$.
Additionally, we show that many estimates of the white 
dwarf space density from different studies are 
consistent with each other, and with our determination here.\\
\end{abstract}


\newpage

\section{Introduction}
One of the main goals of the European Galactic Plane Surveys
(\EGAPS\footnote{EGAPS is the combination of the IPHAS (Drew et al., 2005), 
UVEX (Groot et al., 2009) and the VPHAS+ surveys. Websites: 
http://www.iphas.org, http://www.uvexsurvey.org and http://www.vphasplus.org}) is to
obtain a homogeneous sample of stellar remnants in our Milky Way with
well-understood selection limits. The population of white dwarfs in the
Plane of the Milky Way is relatively unknown due to the effects of
crowding and dust extinction. 
White dwarf space densities and birth rates have mostly been
derived from surveys at Galactic latitudes larger than
$|b|$$>$30\degr, such as the Sloan Digital Sky Survey (\SDSS, York et
al., 2000, Yanny et al., 2009, Eisenstein et al., 2006 and Hu 2007),
the Palomar Green Survey (Green et al., 1986 and Liebert et al., 2005),
the KISO Survey (Wegner et al., 1987 and Limoges $\&$ Bergeron, 2010),
the Kitt Peak-Downes Survey (KPD, Downes, 1986) and the
Anglo-Australian Telescope (AAT) QSO Survey (Boyle et al., 1990). However, as shown in Groot et al. (2009) the
distribution of any Galactic stellar population with absolute magnitude
$M_V$$<$10 is strongly concentrated towards the Galactic Plane in
modern day deep surveys reaching limiting magnitudes of $V$$\sim$22.\\

Within the \EGAPS project, the \UVEX survey images a
10$\times$185 degrees wide band (--5\degr$<$ $b$ $<$+5\degr) centred
on the Galactic equator in the $U,g,r$ and $\he$ bands down to $\sim
21^{st}-22^{nd}$ magnitude using the Wide Field Camera mounted on the
Isaac Newton Telescope on La Palma (Groot et al., 2009). From the
first 211 square degrees of \UVEX data a catalogue of 2\,170 UV-excess
sources was selected in Verbeek et al. (2012a; hereafter V12a). These
UV-excess candidates were identified in the $(U-g)$ versus
$(g-r)$ colour-colour diagram and $g$ versus $(U-g)$ and $g$ versus
$(g-r)$ colour-magnitude diagrams by an automated field-to-field
selection algorithm. Less than $\sim$1$\%$ of these
UV-excess sources were previously known in the literature. A first spectroscopic
follow-up of 131 UV-excess candidates, presented in Verbeek et
al. (2012b; hereafter V12b), shows that 82$\%$ of the UV-excess
catalogue sources are white dwarfs. Other sources in the UV-excess
catalogue are subdwarfs type O and B (sdO/sdB), emission line stars and
QSOs.\\

A determination of the space density and birth rate of hydrogen-line (DA) white dwarfs in the Galactic Plane
based on an observational sample of hot candidate white dwarfs from V12a is presented in this work.
In Sect.\ \ref{sec:theorysample} a theoretical Galactic model
population is constructed to simulate a survey such as \UVEX. In
Sect.\ \ref{sec:observationalsample} the sample of observed candidate white dwarfs
is selected from the UV-excess catalogue, including
an estimate on completeness and homogeneity due to selection effects.
In Sect.\ \ref{sec:method} the method is outlined that is used to
derive the effective temperatures, reddening, and, derived from these,
the distances to the observational sample.
In Sect.\ \ref{sec:results} the method is applied to the observational
sample, considering three sub-samples with slightly different selection biases and
model assumptions. In Sects.\ \ref{sec:spacedensity}
and\ \ref{sec:formationrate}
a space density and birth rate for
hydrogen-line (DA) white dwarfs within a radius of 1 kpc around the Sun are
derived. Taking into account the uncertainties, we give upper and
lower limits on these derived space densities and
birth rates. In Sect.\ \ref{sec:discussion} the impact 
of all assumptions is discussed, the conclusions
are summarized and compared with the results of other surveys.\\

\begin{figure}
\centerline{\epsfig{file=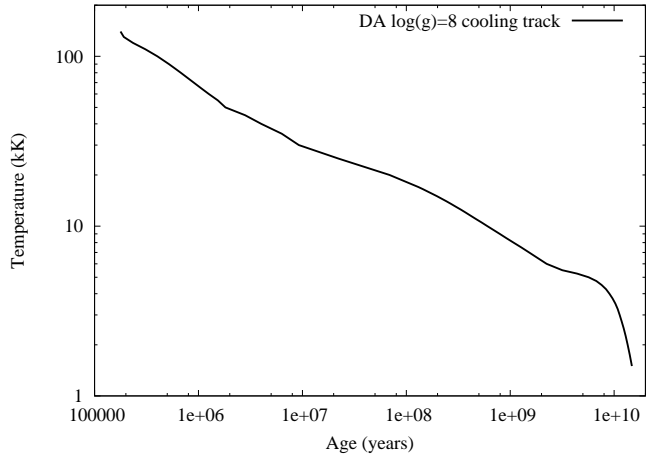,width=9cm,angle=0,clip=}}
\caption{Cooling track for DA white dwarfs with $log(g)$=8 from Wood (1995).
\label{fig:cooling}}
\end{figure}

\begin{figure}
\centerline{\epsfig{file=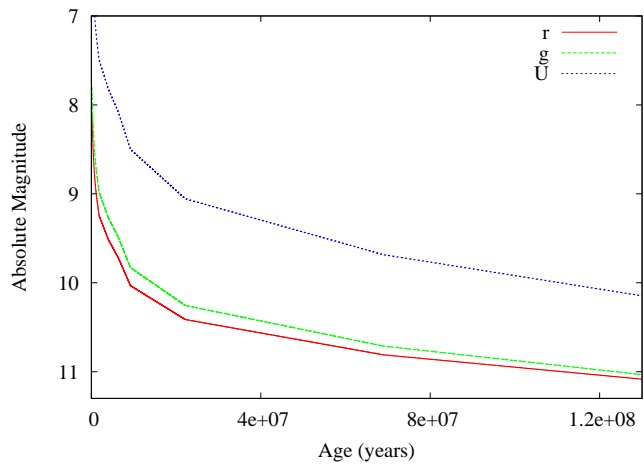,width=9cm,angle=0,clip=}}
\caption{Age versus absolute magnitude for DA white dwarfs with $log(g)$=8 from Holberg $\&$ Bergeron 
(2006) for the UVEX $r$, $g$ and $U$ filter bands.
\label{fig:ageabsolutemag}}
\end{figure}

\begin{figure}
\centerline{\epsfig{file=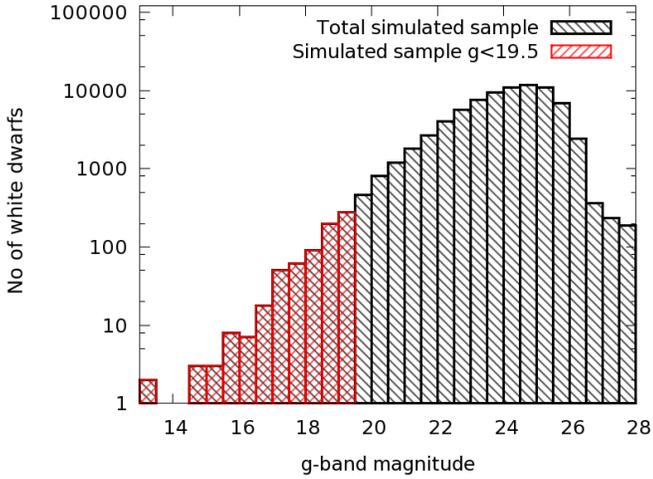,width=9cm,angle=0,clip=}}
\caption{$g$-band magnitude histogram for the modelled white dwarf sample. 
\label{fig:gband}}
\end{figure}

\begin{figure}
\centerline{\epsfig{file=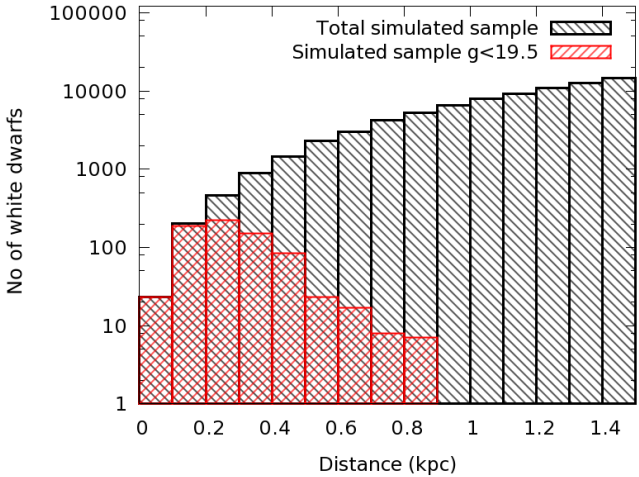,width=9cm,angle=0,clip=}}
\caption{Distance histogram for the modelled white dwarf sample. The number white dwarfs brighter than $g<19.5$ decreases for distances larger than 0.25 kpc while 
the number of white dwarfs in the total simulated sample continues to increase.
\label{fig:distance}}
\end{figure}

\begin{figure}
\centerline{\epsfig{file=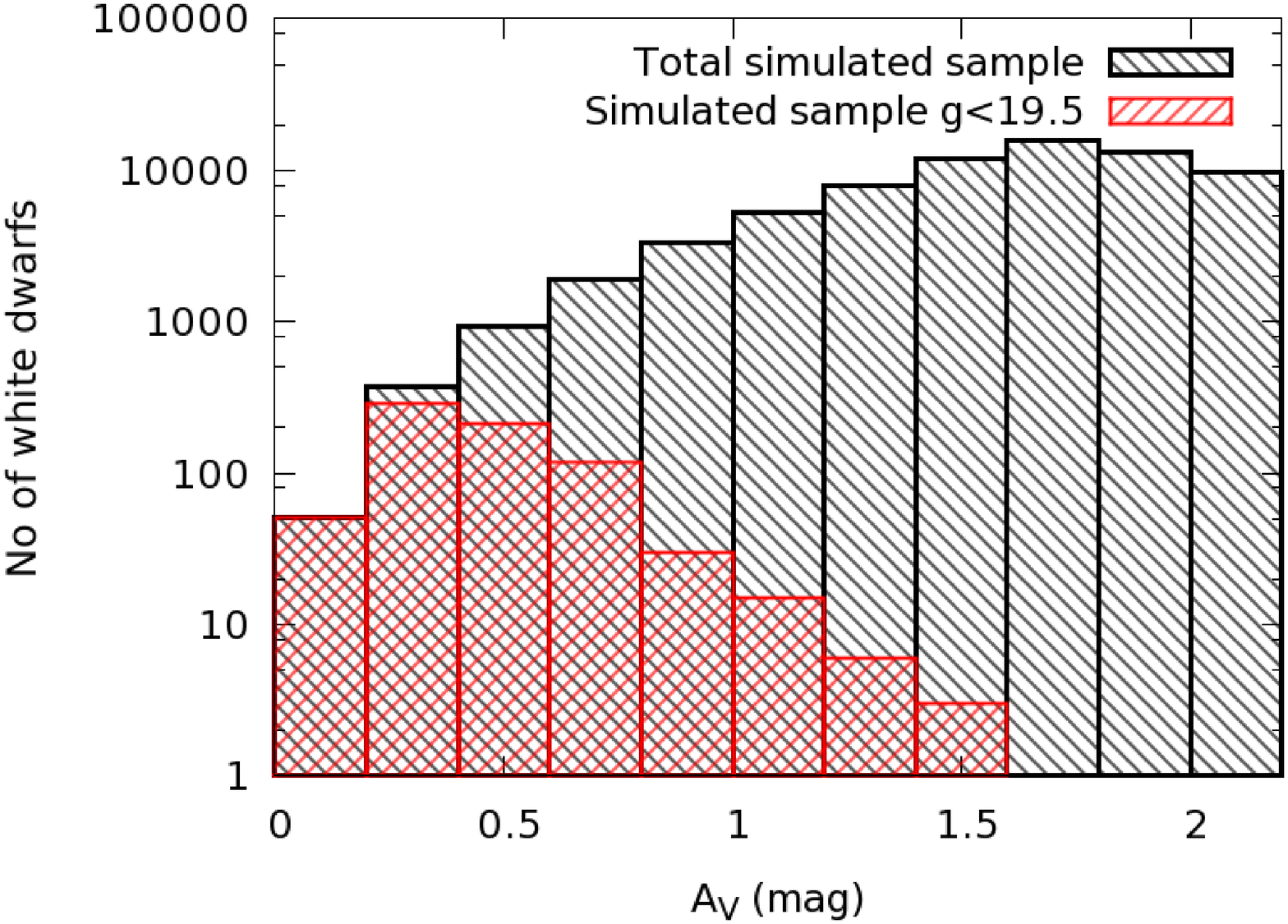,width=9cm,angle=0,clip=}}
\caption{$A_V$ histogram for the modelled white dwarf sample.
\label{fig:av}}
\end{figure}

\begin{figure}
\centerline{\epsfig{file=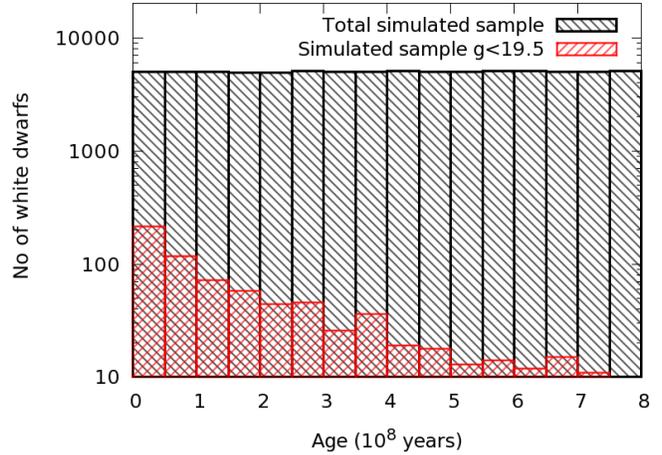,width=9cm,angle=0,clip=}}
\caption{Age histogram for the modelled white dwarf sample. The number of white dwarfs brighter than $g<19.5$ decreases for older systems,
the number of white dwarfs in the total simulated sample is constant for all bins due to the assumption of a constant birth rate.
\label{fig:age}}
\end{figure}

\begin{figure}
\centerline{\epsfig{file=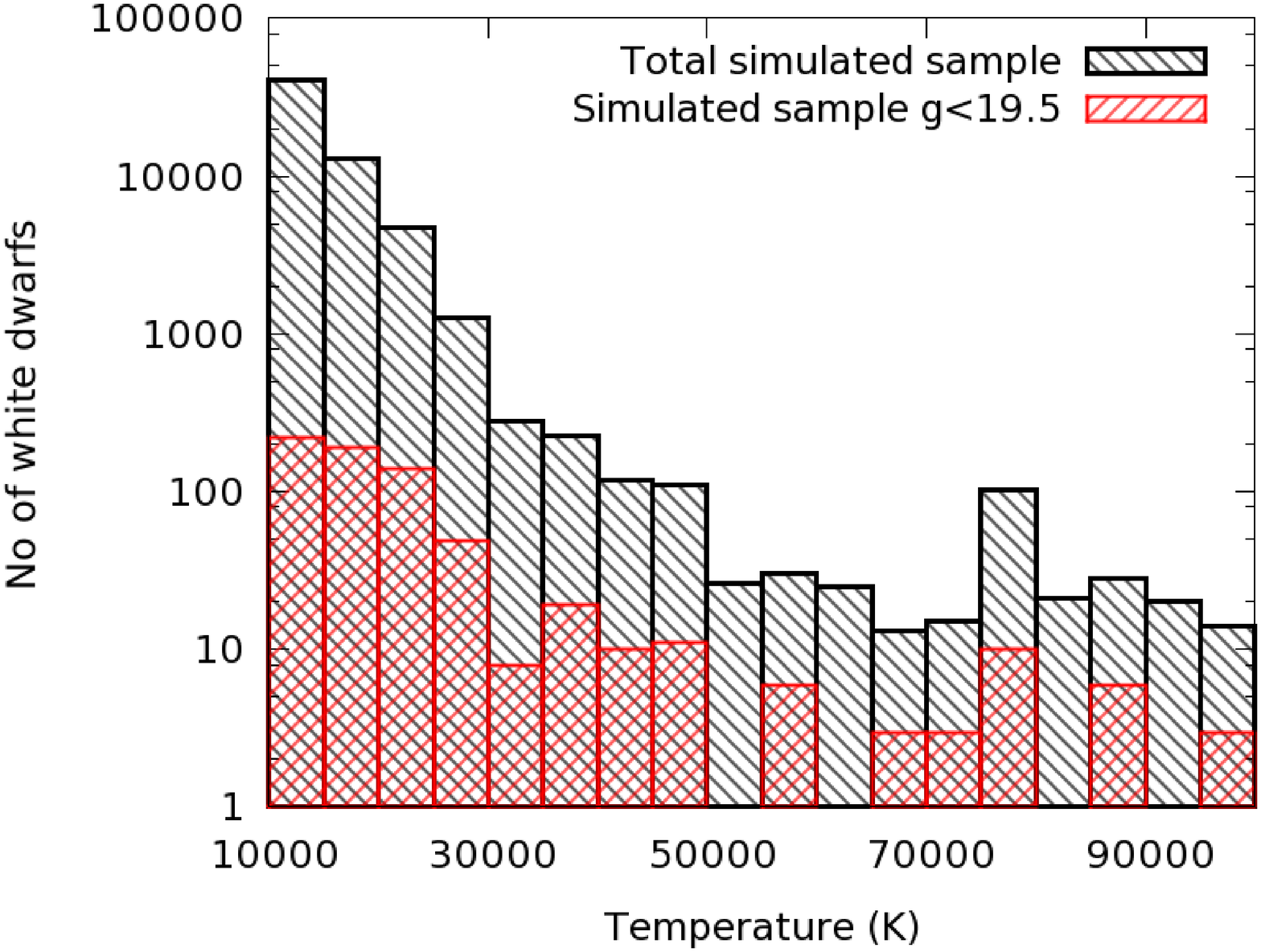,width=9cm,angle=0,clip=}}
\caption{Temperature histogram for the simulated white dwarf sample. 
\label{fig:temperature}}
\end{figure}

\begin{figure}
\centerline{\epsfig{file=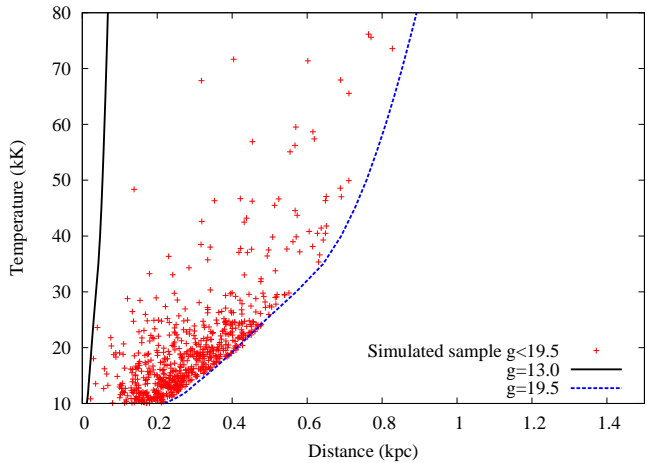,width=9cm,angle=0,clip=}}
\caption{Temperature versus distance for the modelled white dwarf sample (red).
The two lines indicate the survey limits at $g$=13 and $g$=19.5.
\label{fig:distancevstemp}}
\end{figure}

\begin{figure}
\centerline{\epsfig{file=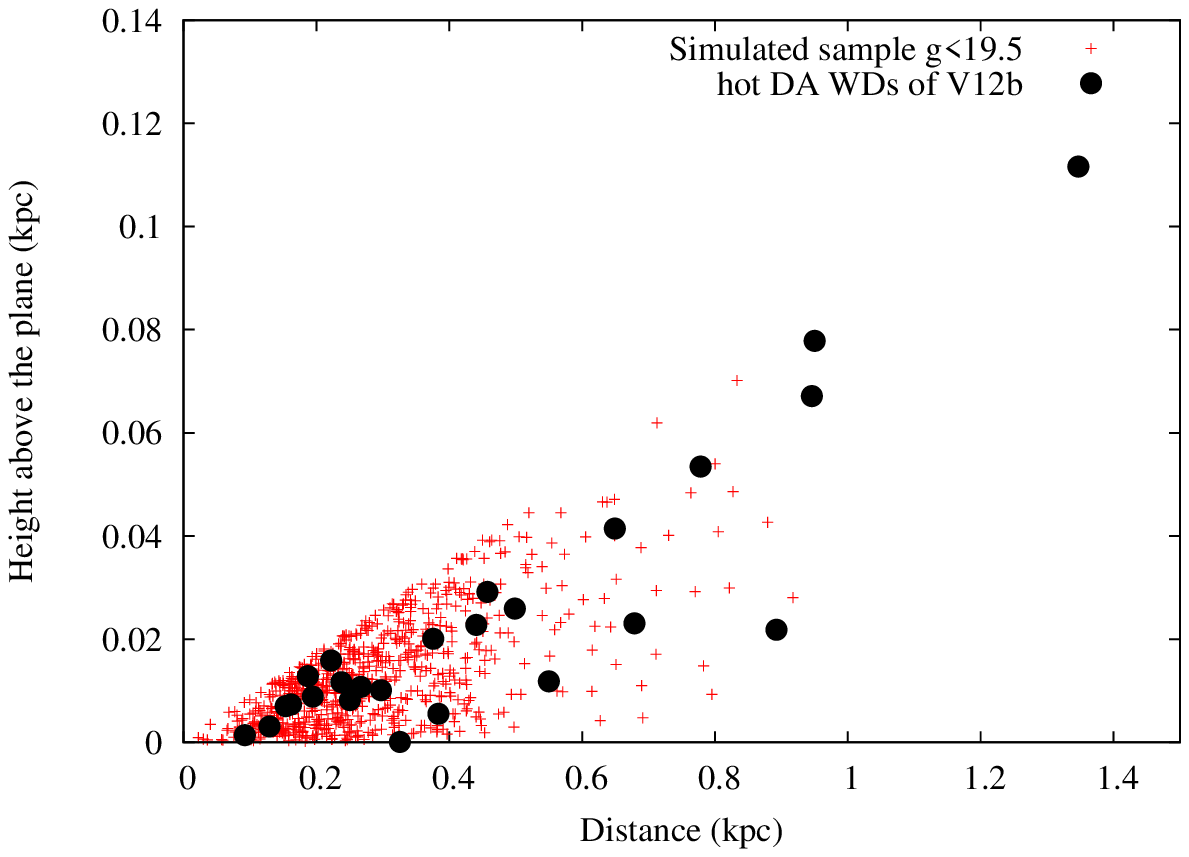,width=9cm,angle=0,clip=}}
\caption{Height above the Galactic Plane versus distance for the modelled white dwarf sample (red) 
and the William Herschel Telescope (WHT) spectroscopically confirmed DA white dwarfs (black) of V12b.
\label{fig:heightvsdistance}}
\end{figure}

\section{A theoretical Galactic Model sample}
\label{sec:theorysample} 
In the determination of the white dwarf space density and birth
rate, the observations will be compared to a Galactic model population
that has been constructed and scaled to emulate the observations. The
model is detailed in Nelemans et al. (2004) and is based on the Galaxy
model according to Boissier $\&$ Prantzos (1999), including a Sandage
(1972) extinction model. 
This model has been presented and used before in
Nelemans et al. (2004), Roelofs, Nelemans \&
Groot (2007) and Groot et al. (2009). 
Note that for the simulated sample in this paper a constant white dwarf birth rate over
the last 8$\times$10$^8$ years is assumed.
Only the spatial distribution of the white dwarfs comes 
from the model as described in Nelemans et al. (2004).\\

A theoretical sample of 8$\times$10$^4$ DA white dwarfs is generated
by a random distribution over ages 0-8$\times$10$^8$ years
and a model-weighted position in the Galaxy with a Galactic longitude
spread of --180\degr$<$$l$$<$+180\degr, a Galactic latitude spread
--5\degr$<$$b$$<$+5\degr (the limits of the \EGAPS surveys), a
distance $d$$<$1.5 kpc and an extinction $A_V$ appropriate to their
distance and Galactic location.\\

Using the DA white dwarf cooling models by Wood (1995)
the adopted age distribution ($<$8$\times$10$^8$ years) translates into a minimal
temperature of T$_{\rm eff}$$>$9\,000 K.
This cut in temperature is required due to the uncertainties and incompleteness of colder white dwarfs 
in the observed sample (see Sect.\ \ref{sec:observationalsample}).
The temperature of the white dwarf is related to an 
absolute magnitude ($M$) for each filter band (Figs.\ \ref{fig:cooling}
and\ \ref{fig:ageabsolutemag}).
For each object, the absolute magnitude in the Vega system in the
\UVEX bands ($M_{U}$, $M_{g}$, $M_{r}$) has been calculated using the colour calculations 
of Holberg \& Bergeron\footnote{http://www.astro.umontreal.ca/$\sim$bergeron/CoolingModels} (2006), 
Kowalski \& Saumon (2006), Tremblay et al. (2011) and Bergeron et al. (2011),
assuming a surface gravity $\log g$ = 8.0, and the \UVEX filter passbands presented in Groot et al. (2009).
The magnitudes are converted to the Vega system using the AB offsets $U$=-0.927, $g$=0.103, $r$=-0.164 of
Gonz\'{a}lez-Solares et al., 2008 and Hewett et al., 2006.
These values need to be added to the AB magnitudes to convert them to the Vega system.
Reddened apparent magnitudes and Vega colours for each white dwarf were calculated using
$A_\lambda/A_V$=1.66, 1.16, 0.84 for the $U$-, $g$- and $r$-bands
respectively. To emulate the observational sample of
Sect.\ \ref{sec:observationalsample}, only white dwarfs with $g$$<$19.5
were selected, keeping 723 systems out of the original model
sample. The reason of the magnitude cut at $g$$<$19.5 is to warrant 
the completeness of the observational sample. Going deeper clearly showed a down-turn in
the number of systems in the observational sample, indicative of loss
of completeness.\\

The characteristics of the simulated white dwarf sample are shown in
Figs.\ \ref{fig:gband}-\ref{fig:heightvsdistance}.
Figure\ \ref{fig:gband} shows that the number of white dwarfs
keeps on increasing for magnitudes $g$$>$19.5 and only turns over around $g$$\sim$25 due to the combined
effects of a minimum temperature and a limited volume in the model. 
White dwarfs detectable in a survey such as \UVEX are only a tip of the iceberg
compared to the total population, even within a limited distance
of $\sim$1 kpc. No white dwarfs in the sample are brighter than
$g$$\sim$13.3 and all white dwarfs brighter than $g$$<$19.5 are within
$d$$<$0.92 kpc (Fig.\ \ref{fig:distance}). The observable sample is complete to a
distance of $\sim$0.2kpc and all systems have extinctions
smaller than $A_V$$<$1.7 (Figs.\ \ref{fig:distance} and\ \ref{fig:av}).
$A_V$ peaks between 0.3 and 0.4 for the white dwarfs brighter than $g<19.5$ while it peaks between
1.7 and 1.8 for the total simulated white dwarf sample.\\

The age distribution is shown in Fig.\ \ref{fig:age}, and the
corresponding white dwarf temperatures in
Fig.\ \ref{fig:temperature}. The fraction of hot, young white dwarfs is
larger than the fraction of older, colder white dwarfs.
The number of white dwarfs in the complete simulated sample 
varies with different Galactic latitude and longitude.
While the number of white dwarfs in the total simulated sample is
higher at Galactic latitude $b$=0\degr\ and Galactic longitude
$l$=0\degr, the number in the $g$$<$19.5 sample is constant over
Galactic latitude and Galactic longitude.
The white dwarfs brighter than $g$$<$19.5 are equally
distributed over different Galactic latitudes and Galactic longitudes
due to the limited distance probed in this first, relatively shallow sample. 
The distributions over distance, temperature and height above
the Galactic Plane of the numerical, observable sample are shown
as red points in Figs.\ \ref{fig:distancevstemp} and\ \ref{fig:heightvsdistance}. 
The lines in Fig.\ \ref{fig:distancevstemp} show for the chosen upper
and lower magnitude limits in the observable out to which distance a
survey such as \UVEX is sensitive as a function of temperature.\\

\begin{figure*}
\centerline{\epsfig{file=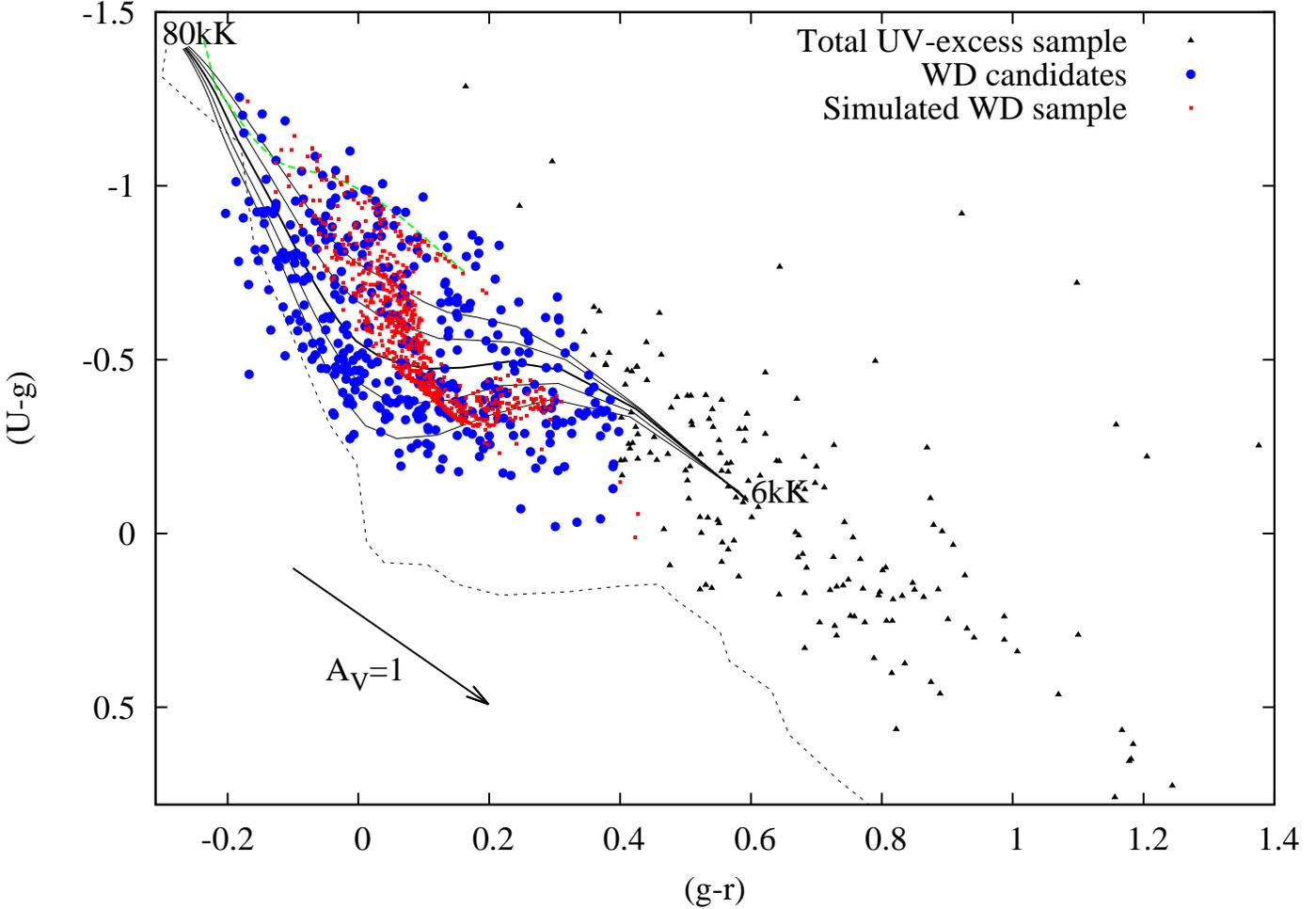,width=19cm,angle=0,clip=}}
\caption{Colour-colour diagram with the UV-excess sources of V12a
  brighter than g$<$19.5 (black triangles), the white dwarf candidates in the
  UV-excess catalogue (blue circles), and the simulated white dwarf sample
  brighter than g$<$19.5 (red squares). Overplotted are the simulated colours
  of unreddened main-sequence stars (dashed black) and the simulated
  colours of unreddened Koester DA (solid black) and DB (dashed green) white
  dwarfs of V12a. The DA white dwarf colours are shown for different
  $log\,g$=7.0, 7.5, 8.0, 8.5, 9.0, where the upper line is
  $log\,g$=9.0. The DA white dwarf colours cover the range 80kK$>T_{\rm eff}>$6kK
  No photometric error bars are plotted for
  clarity, photometric errors range from 0.002 mag at $g$=16 to $<$0.025 mag at
  $g$=19.5. The simulated DA white dwarfs are distributed over the Galaxy and reddened to their
  Galactic position. The simulated sources around the unreddened DB track (dashed green) 
  are reddened hot DA white dwarfs. The vector ($A_{V}$=1) shows the direction of 
  the reddening, its length is equal to $A_{V}$=1.
\label{fig:ccdcompare}}
\end{figure*}

\begin{figure*}
\centerline{\epsfig{file=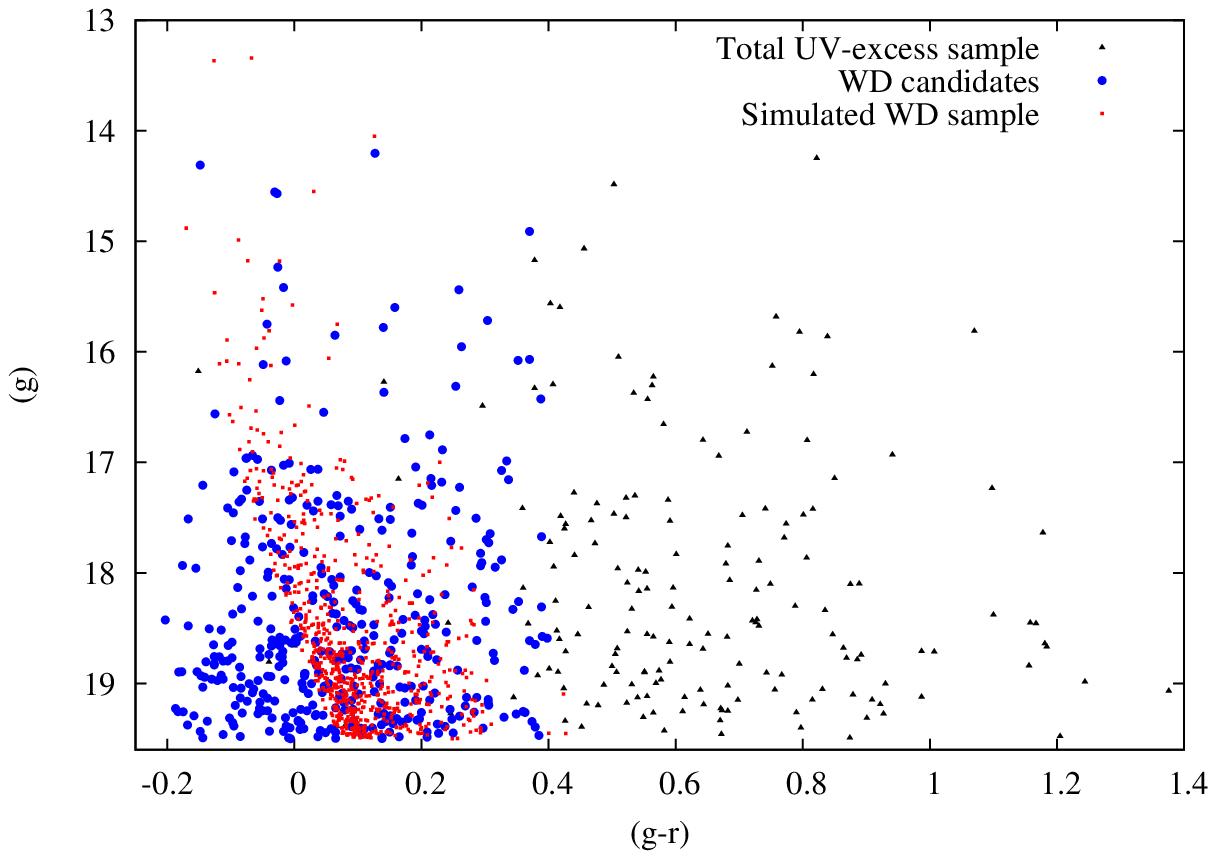,width=19cm,angle=0,clip=}}
\caption{Colour-magnitude diagram with the UV-excess sources of V12a
  brighter than g$<$19.5 (black triangles), the white dwarf candidates in the
  UV-excess catalogue (blue circles), and the simulated white dwarf sample
  brighter than g$<$19.5 (red squares). No photometric error bars are plotted for
  clarity, photometric errors range from 0.002 mag at $g$=16 to $<$0.025 mag at
  $g$=19.5.
\label{fig:cmd}}
\end{figure*}

\section{The Observational white dwarf sample}
\label{sec:observationalsample}
The UV-excess catalogue of V12a, selected from the first 211 square
degrees of \UVEX data 
contains 2\,170 UV-excess sources. 
In the colour-colour and colour-magnitude diagrams an automated algorithm 
selects blue outliers relative to other stars in the same field.
We have used recalibrated \UVEX
photometry to correct for the time-variable $U$-band calibration as
noted in Greiss et al. (2012). The recalibrated \UVEX data are
explained in Appendix\ \ref{app:recalibrated}.\\

Since we are interested in a complete sample of white dwarfs
with minimal pollution we select all sources with $g$$<$19.5 and
$(g-r)$$<$0.4. 
The distribution of the observational sample for magnitudes fainter than $g$$>$19.5 clearly
showed a down-turn indicative of loss of completeness, therefore a magnitude cut at $g$=19.5 is applied.
The colour cut at $(g-r)$$<$0.4, which corresponds to the colour of 
unreddened DA white dwarfs with $T_{\rm eff}$$\sim$7\,000K,
is applied since all DA white dwarfs in V12b have $(g-r)$$<$0.4 (Figs.\,1 and 2 of V12b).
From spectroscopic follow-up of V12b it is known that the observational sample 
with $(g-r)$$<$0.4 is dominated by DA white dwarfs, and spectroscopy of
the ``subdwarf sample'' of V12a shows that the UV-excess catalogue is 
complete for white dwarfs.
The effects of the pollution of the observational sample 
are corrected in Sect.\ \ref{sec:completeness}.
Additionally, sources more than 0.1 magnitude above the reddened 
white dwarf locus in the $(g-r)$ vs. $(U-g)$ colour-colour diagram
are not taken into account since they are more than 0.1 magnitude above
the reddened hottest white dwarf model.\\

These cuts result in an sample of 360 observed candidate white
dwarfs, which will be used as the basis of the space density
and birth rate calculations in this paper. The observational white dwarf sample is
shown in the colour-colour and colour-magnitude diagrams of
Figs.\ \ref{fig:ccdcompare} and\ \ref{fig:cmd}, overplotted by the
simulated sample of Sect.\ \ref{sec:theorysample}. 
The objects in the simulated sample have colours similar to reddened 
synthetic colours shifted by a defined amount of reddening, 
determined by the Galactic position of the objects in the theoretical sample.
The simulated white dwarf sample and the observed white dwarf sample have different colours since the
observed sample has a photometric scatter and an uncertainty on the absolute calibration. 
Additionally, from V12b it is known that the observed sample contains some non-DA sources, 
such as DB white dwarfs. We correct for these non-DAs in Sect.\ \ref{sec:completeness}.\\

\begin{figure}
\centerline{\epsfig{file=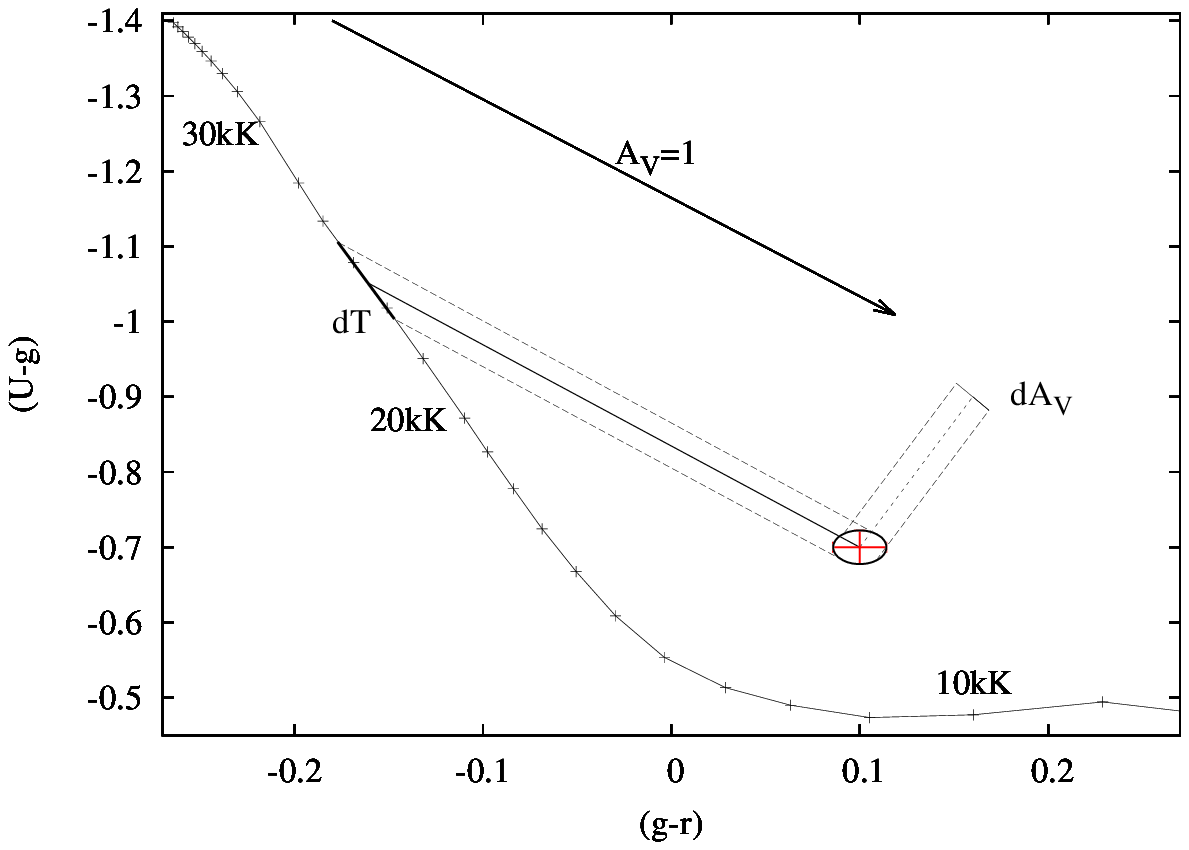,width=9cm,angle=0,clip=}}
\caption{The temperature and reddening of each white dwarf in the
  observational sample are determined from the location in the
  colour-colour diagram. The temperature is determined by tracing the
  reddening vector (thick line) back to the unreddened synthetic
  $log\,g$=8.0 DA white dwarfs model colours of V12a. The reddening
  corresponds with the length of the reddening vector. The error on
  the determined temperature and reddening are related the photometric
  error and the error on the temperature depends on the position of
  the source in the colour-colour diagram.
\label{fig:vectormethod}}
\end{figure}

\begin{figure}
\centerline{\epsfig{file=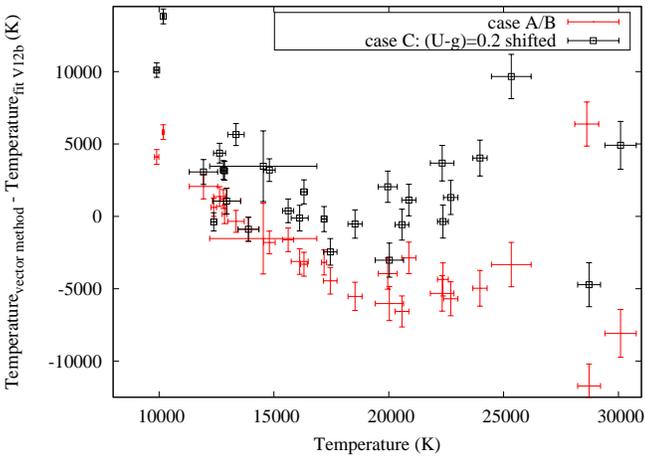,width=9cm,angle=0,clip=}}
\caption{Difference between the temperature determined from the
  photometry by the vector method and the temperature determined
  through line profile fitting in V12b. When the vector method is
  applied to the original UVEX data (cases A and B) the difference in
  temperature increases for hotter white dwarfs with a clear trend.
  For the UVEX with a shift of $(U-g)$=0.2 magnitude (case C) the
  difference in temperature is more scattered around zero without this
  trend. There are two more hot white dwarfs at $35kK<T_{\rm eff}<40kK$ 
  that have a temperature difference larger than 15kK.
\label{fig:howgoodistempmethod}}
\end{figure}

\section{Space densities and birth rates: Method} 
\label{sec:method}
The method to derive the space density and birth rate is first
to calculate for each system in the observational sample its
current temperature and distance, and then to scale and compare
the distribution of the observed sample to the simulated,
numerical sample. Herein we adjust the space density of the
numerical sample to match the observed number of systems. The test
we perform here is therefore how well the observed population
resembles a simulated, numerical sample. In
Sect.\ \ref{sec:discussion} we will discuss the validity of our
assumptions in constructing the numerical, observable sample.\\

To derive temperatures and distances the observed position of a source
in the $(g-r)$ and $(U-g)$ diagram is compared with a grid of reddened
model colours, based on the hydrogen dominated white dwarf atmosphere
models of Koester et al. (2001) for temperatures in the range 
6\,000$\leq T_{\rm eff}$$\leq$80\,000 K. We assume a fixed surface gravity
of $\log g$ = 8.0, as this is the median value found in the
spectroscopic fitting of a representative sample of the white dwarf
systems in V12b (Fig.\,5 of V12b, and e.g. Fig.\,5 of Vennes et al.,
1997; Fig.\,9 of Eisenstein et al., 2006). 
The impact of this assumption of a fixed surface gravity of $\log g$ = 8.0
is discussed in Sect.\ \ref{sec:discussion}.
The reddening grid was calculated at $\Delta E(B-V)$=0.1 intervals. 
The best-fitting value for an individual system is taken as the grid point with the smallest
distance to the observed value (see Fig.\ \ref{fig:vectormethod}).
Error estimates on the fit values
are obtained by projecting the 1$\sigma$ photometric errors on to the
grid axes, often resulting in asymmetric errors in temperature. The
distance to a source is derived from the combination of the
observed $g$-band magnitude, the model absolute magnitude
corresponding to the surface gravity and the derived temperature, 
including the reddening value derived from the fit.\\

As the synthetic colour tracks of white dwarfs display a distinct
`hook' in the colour-colour plane at $T_{\rm eff}$=10\,000K (Fig.\,10), due to the
strength of the Balmer jump, 
a highly reddened object may have a dual possible solution: a high
temperature/high reddening, or a low temperature/low
reddening solution. The numerical model shows that in most of
the cases where a dual solution exists, the preferred one is the
hot solution. In observational samples cool white dwarfs below $T_{\rm eff}$$<$10\,000K
are more rare (Fig.\,5 V12b, Eisenstein et al., 2006 and Finley et al., 1997).
Therefore it is assumed that in all cases the hot
solution is correct. The impact of this assumption will be discussed
in Sect.\ \ref{sec:discussion}.\\

Fig.\ \ref{fig:ccdcompare} shows the colour-colour diagram of the selected
\UVEX sample alongside theoretical tracks of cooling white
dwarfs of various surface gravity ($\log g$ = 7.0 - 9.0, from bottom
to top). The spectroscopic analysis of V12b (Fig.\,5) shows that the vast
majority of the DA white dwarfs in the \UVEX sample have $\log g$$\sim$8, 
and should therefore lie at the $\log g$=8 line. However, a substantial
number of systems in Fig.\,7 of V12b lie below the line, either
due to their own photometric error, the scatter in the absolute calibration in the
$U$-band magnitude, the lack of a global photometric calibration in
the \UVEX survey, or a combination of all three. To investigate the
effect of this scatter on the determination of the space densities and
birth rates three separate samples are defined and analysed.\\

\begin{itemize}
\item {\it Sample A} (only systems above $\log\,g$=8.0 line):
In sample A, 84 white dwarfs which are located far under/left of the unreddened synthetic $\log\,g$=8.0 
colour track of Fig.\ \ref{fig:ccdcompare} are not taken into account. 
To allow for some intrinsic photometric scatter all systems that lie closer than 0.1
magnitude left of the unreddened synthetic $\log\,g$=8.0 colour track are included, 
automatically have reddening $E(B-V)$=0, and are assigned the temperature of the grid point on the track 
most closely located to the measurement. Sample A contains 276 white dwarf systems.\\

\item {\it Sample B} (all systems):
In sample B all candidate white dwarfs shown in Figs.\ \ref{fig:ccdcompare}
and\ \ref{fig:cmd} are included.
Temperatures and reddening vectors are compared to $\log g$=8.0 models only. 
For systems below the $\log g$=8.0 line,
temperatures and reddenings are assumed to be those of the
closest grid point on the $\log g$=8.0 line. This sample will
correctly include the number of systems present, but will overestimate
the number of systems at very low reddenings. Sample B contains
360 systems in our footprint.\\

\item {\it Sample C} (shifted $U$-band): In sample C a
shift of $(U-g)$=--0.2 is applied to all systems in the observational sample. 
This brings the vast majority of systems above the $\log g$=8.0 line. 
The magnitude of the shift is the maximum scatter
observed in the $U$-band calibration and will therefore in general
overestimate the actual calibration uncertainty. 
A consequence of the $(U-g)$ shift is that each 
white dwarf will get a different colour, and so a different $T_{\rm eff}$, $E(B-V)$ and distance.
Sources that are more than 0.1 magnitude above the white dwarf grid 
after the colour shift are not taken into account. 
This sample excludes a fraction of systems and will lead to 
an overestimate of the number of hot, distant systems. 
Sample C contains 303 systems in our footprint.\\
\end{itemize}

Note that samples A and C do not give the best exact 
value of the space density and birth rate, 
but are presented to show the effect of a colour shift or cut $\log g$$>$8.
To obtain a feel for the accuracy in temperature and reddening from
the photometric method, the effective temperature ($T_{\rm eff}$),
surface gravity ($\log g$) and reddening of the 20 \UVEX white dwarfs
with $T_{\rm eff}$$>$20\,000K in V12b (Table\,2, Fig.\,5)
were compared with the photometric method, with and without
applying any shifts, and the results are shown in
Fig.\ \ref{fig:howgoodistempmethod}. For samples A and B,
the difference in effective temperature found by the photometric
method and determined through line profile fitting 
of V12b appears to increase for hotter white dwarfs. 
This trend is less clear for sample C, see Fig.\ \ref{fig:howgoodistempmethod}.\\

In the photometric method the error on the temperature depends on the $(g-r)$ and
$(U-g)$ colours of the white dwarf. The errors on the temperature and
reddening due to the method are between $\Delta T_{\rm eff}$=2\,000K
for white dwarfs of $T_{\rm eff}$=25\,000 K and $\Delta T_{\rm eff}$=15\,000K 
for white dwarfs of $T_{\rm eff}$=60\,000 K and
$\Delta E(B-V)$$\sim$0.03 for sources with a photometric error of $\Delta
g$=0.01 mag. However, the uncertainty in temperature and reddening is
larger because of the lack of a global photometric calibration.\\

Photometric distances ($d$) to all systems in the observed
samples were calculated using 
$$d=0.01 \times 10^{0.2(m_{g}-M_{g}-A_{g})} (kpc)$$, 
where $m_g$ and $M_g$ are the observed and absolute $g$-band magnitude and $A_{g}$ is the
extinction in the $g$-band. The absolute magnitudes of Holberg \& Bergeron (2006) are used, 
assuming $\log g$=8.0 for all white dwarfs.\\

\begin{figure}
\centerline{\epsfig{file=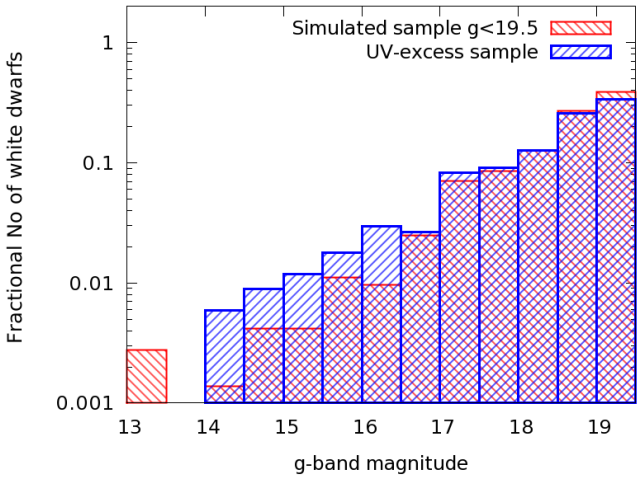,width=9cm,angle=0,clip=}}
\caption{Histogram of g-band magnitudes of the UV-excess white dwarf
  candidates of sample B and the simulated white dwarf
  sample.
\label{fig:wdhistogramgband}}
\end{figure}

\begin{figure}
\centerline{\epsfig{file=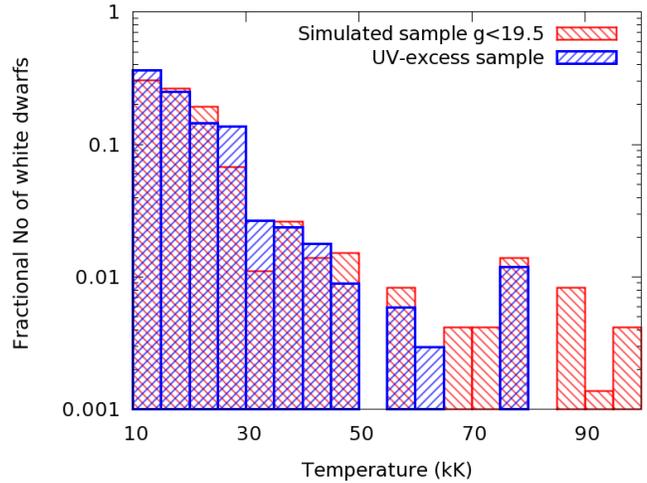,width=9cm,angle=0,clip=}}
\caption{Temperature histogram of the UV-excess candidate white dwarfs
  of sample B and the simulated white dwarf sample. 
\label{fig:wdhisttemperature}}
\end{figure}

\begin{figure}
\centerline{\epsfig{file=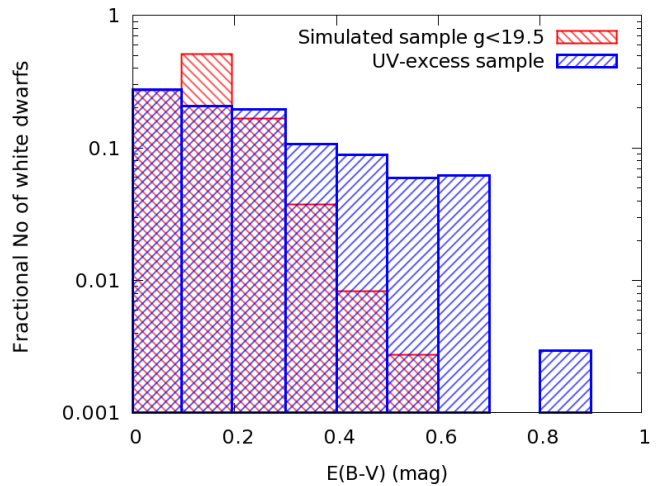,width=9cm,angle=0,clip=}}
\caption{Histogram of $E(B-V)$ of the UV-excess candidate white dwarfs of sample B and the simulated white dwarf sample.
\label{fig:wdhistogramav}}
\end{figure}

\begin{figure}
\centerline{\epsfig{file=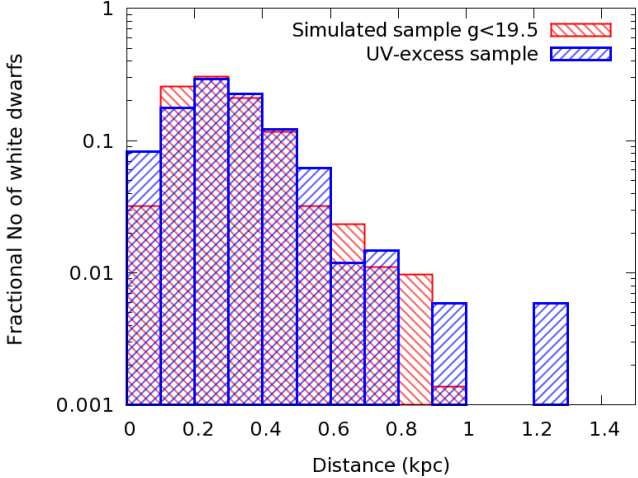,width=9cm,angle=0,clip=}}
\caption{Distance histogram of the UV-excess candidate white dwarfs of sample B and the simulated white dwarf sample. 
\label{fig:wdhistdistance}}
\end{figure}

\begin{figure}
\centerline{\epsfig{file=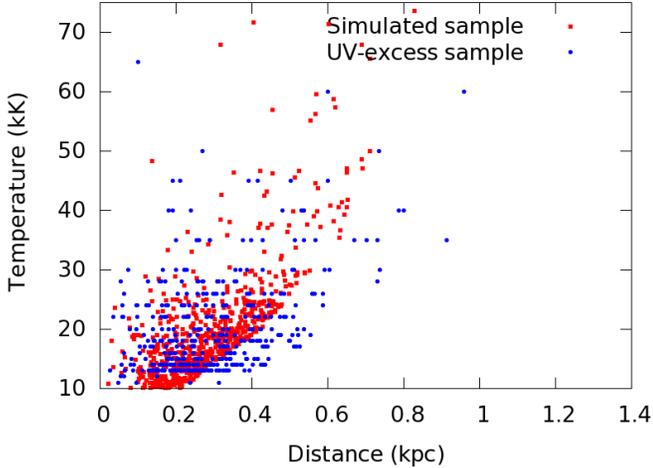,width=9cm,angle=0,clip=}}
\caption{Temperature versus distance of the UV-excess candidate white dwarfs 
(sample B) from UVEX (blue) and the simulated sample (red).
\label{fig:wdtempvsdistance}}
\end{figure}

\begin{figure}
\centerline{\epsfig{file=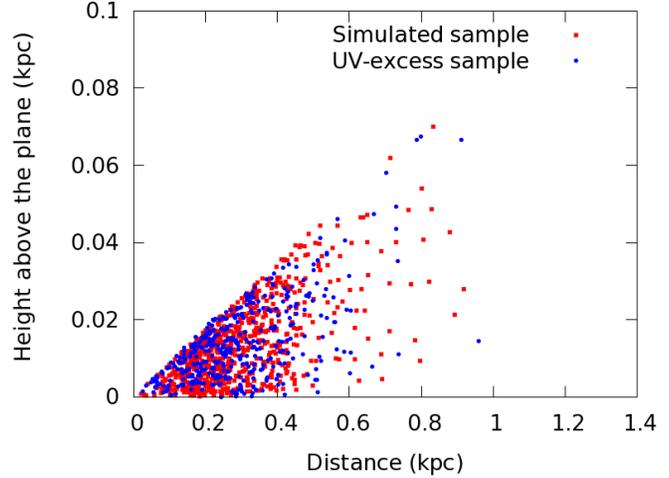,width=9cm,angle=0,clip=}}
\caption{Height above the plane versus distance of the UV-excess candidate white dwarfs
(sample B) from UVEX (blue) and the simulated sample (red).
\label{fig:wdheightvsdist}}
\end{figure}

\section{Results}
\label{sec:results}
For illustration purposes only the distribution comparisons for sample B are 
shown here (Figs.\ \ref{fig:wdhistogramgband} to\ \ref{fig:wdhistdistance}).
The histograms of all three samples (A-C) are shown in Figs.\ \ref{fig:magn3samples}
to\ \ref{fig:dist3samples} of Appendix\ \ref{app:histograms}.
For the magnitude, reddening, distance and temperature distributions, 
sample B is most consistent with the simulated sample.
Because sample B is most complete and has the best Kolomogorov-Smirnov (KS) results, this sample will
be emphasized in the next sections.\\

\begin{table*}
\caption[]{KS tests between the derived, observed,
  and numerical distributions for samples A-C. \label{tab:KStest} }
\begin{tabular}{ | l | c | c | c | }
    \hline
Distribution & Sample A & Sample B & Sample C
\\
& $(D,p)$ & ($(D,p)$ & $(D,p)$\\
 \hline
Magnitude   & (0.09,0.07)& (0.08,0.11) & (0.09,0.09) \\
Reddening  & (0.40,4.0$\times$10$^{-26}$)& (0.30,1.2$\times$10$^{-18}$) & (0.41,9.6$\times$10$^{-30}$) \\
Distance   & (0.07,0.25)& (0.06,0.49) & (0.22,1.2$\times$10$^{-8}$) \\ 
Temperature & (0.20,6.2$\times$10$^{-7}$)& (0.17,3.0$\times$10$^{-6}$) & (0.36,1.1$\times$10$^{-22}$) \\ \hline
\end{tabular}
\end{table*}

To test whether the distributions in temperature, distance and reddening between the
observational and numerical sample are consistent with each other, 
a KS test was performed on the cumulative
distributions in magnitude, reddening, distance and temperature,
with limiting values of 13$<$$g$$<$19.5, 0$<$$E(B-V)$$<$0.7, 
0$<$$d$(kpc)$<$1.0 and 10\,000$<$$T$(K)$<$ 80\,000. 
The results of the KS-test are summarized in Table\ \ref{tab:KStest}.
The D-value is the maximum distance between the cumulative distributions and 
the p-value is the probability that the observational and numerical samples are
the same distributions.
If the D-value is small or the p-value is high, the hypothesis cannot be rejected that the 
distributions of the numerical and observational samples are the same.\\

Our main conclusion from Table\ \ref{tab:KStest} and the distributions shown in
Figs.\ \ref{fig:wdhistogramgband} to\ \ref{fig:wdhistdistance} is 
that the numerical sample reproduces
the reconstructed observational samples reasonably accurate, except for the
reddening and temperature, where, for sample B, the reddening gradient is too
shallow. There are not enough observed systems at low reddening, and too many
at high reddening. Note however, that the model reddening is a
very simple Sandage-type relation and can therefore easily
underestimate the amount of reddening in the local volume.
\\

Figs.\ \ref{fig:wdtempvsdistance} and\ \ref{fig:wdheightvsdist} show
the main similarities and differences between the observed white dwarf
sample and the theoretical $g$$<$19.5 sample. 
In Fig.\ \ref{fig:wdtempvsdistance} there is a clear lower limit in the 
distance-temperature distribution due to a linear relation between
the distance and reddening for the theoretical white dwarfs, 
while there are candidate white dwarf in the observational sample that have little reddening at a large distance 
or strong reddening at a small distance as a result of the method in Sect.\ \ref{sec:method}.
Note that due to the method of Sect.\ \ref{sec:method} the results of $T_{\rm eff}$ and $E(B-V)$ are strongly correlated.
All white dwarfs have a height above the plane smaller than 0.07 kpc (Fig.\ \ref{fig:wdheightvsdist}),
which is a consequence of the \UVEX Galactic latitude limit $|b|$$<$5\degr,
and all white dwarfs are within a distance of 1.0 kpc due to the brightness limit of \UVEX.\\

The vector method finds several solutions at $T_{\rm eff}$=80kK since that 
is the hottest model of the white dwarf grid, see Fig.\ \ref{fig:wdhisttemperature}.
A small fraction of these sources might be photometrically scattered DA white dwarfs,
white dwarfs hotter than $\Delta T_{\rm eff}$$>$80\,000K,
non-DA white dwarfs (DB, DA+dM) or subdwarfs.
As shown in V12b and follow-up spectroscopy of the comparable region
in the Sloan Digital Sky Survey (Rau et al., 2010; Carter et al., 2012, and V12b), 
the majority of these objects are helium-line (DB) white dwarfs, subdwarfs, DA+dM stars and
Cataclysmic Variables.
These hottest solutions also induce the peaks at 0.8$<$$E(B-V)$$<$0.9 and
1.2$<$$d(kpc)$$<$1.3 in Figs.\ \ref{fig:wdhistogramav}
and\ \ref{fig:wdhistdistance}.
\\

The maximum reddening of $E(B-V)$=0.7 in Fig.\ \ref{fig:wdhistogramav} corresponds
with a maximum extinction of $A_{V}$=2.2 using $R_V$=3.1 and is in
agreement with Fig.\,8 of V12b. The difference between the theoretical
and observational sample for reddening smaller than $E(B-V)$$<$0.2 might be due to the method to
determine the temperature and reddening of the UV-excess sources below
the unreddened white dwarf track. The method assumes the nearest
grid point for these sources while these sources might be slightly
more reddened in reality.\\

\section{Completeness of the observed sample}
\label{sec:completeness}
Before space densities and birth rates can be derived by scaling the
observed sample to the numerical sample, a number of corrections to
the observed sample need to be applied:\\

\subsection{Non-DA white dwarf selection}
\label{sec:nonda}
In the spectroscopic follow-up of the UV-excess catalogue presented in
V12b, it was concluded that 67$\%$ of the sources with $g$$<$19.5 and $(g-r)$$<$0.4
were indeed DA white dwarfs (Fig.\,1 and Fig.\,2 of V12a). Fifteen percent was classified as white dwarfs
of other types (DB, DAB, DC, DZ, DA+dM, DAe) and 18$\%$ were non-white dwarfs
(Cataclysmic Variables, Be stars, sdO/sdB stars).
For a correct derivation of the space number density
and birth we correct for the fraction of genuine DA white dwarfs 
and take the 67$\%$ into account. This is the largest
correction made to the observed numbers.\\

\subsection{Non-selection of DA white dwarfs}
\label{sec:nonselection}
From spectroscopic follow-up of the `subdwarf sample' (see V12a) it is concluded 
that the method described in V12a selects all
observable white dwarfs, so the observational sample is almost complete in its selection of
white dwarfs with temperatures $T_{\rm eff}$$>$10\,000K ($M_{V}$$<$12.2). 
In both the theoretical and observational samples, only white 
dwarfs hotter than $T_{\rm eff}$$>$10\,000K
are taken into account, due to the distinct `hook' in the synthetic 
colours of white dwarfs in the $(U-g)$ vs. $(g-r)$ colour-colour diagram
The brightest UV-excess candidates with $g$$<$16 and
$(g-r)$$<$0.4 have a chance not to be selected, see Fig.\,14 of
V12a. The sources brighter than $g$$<$16 are a small fraction of only
2$\%$ of the theoretical sample and 3$\%$ of the observational
sample. Additionally, in the simulated white dwarf sample there are 3 sources with
$(g-r)$$>$0.4, so some reddened white dwarfs could be missed in the
observational sample because they are at $(g-r)$$>$0.4.
For the derivation of the space number density
and birth, these effects are not taken into account, 
since both contributions are negligible.\\

\begin{figure}
\centerline{\epsfig{file=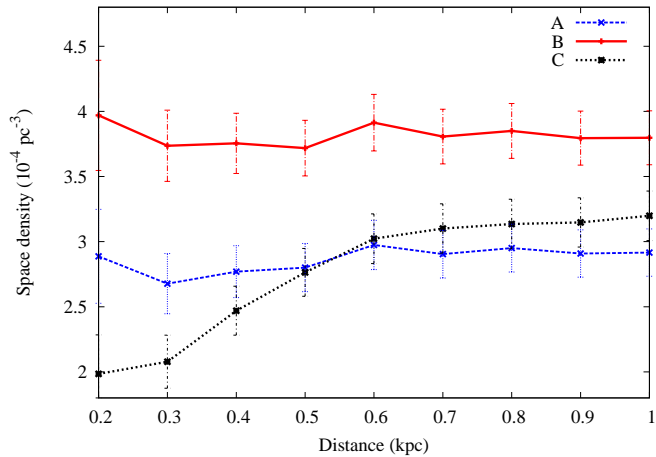,width=9cm,angle=0,clip=}}
\caption{Space density versus distance for the three different samples (A, B, C), 
demonstrating that the space densities derived only slightly depend on the volume.
Here the error bars indicate the number of white dwarfs ($N$) used to derive each space density
and are calculated as $1/\sqrt(N)$. 
\label{fig:rhodistance}}
\end{figure}

\begin{table*}
\caption[]{Space densities for DA white dwarfs with $T_{\rm eff}>$10\,000K and birth rates from \UVEX. \label{tab:sdbruvex} }
\centering
{\small
\begin{tabular}{ | l | c | c | r | }
    \hline
Case     &  Space density $(10^{-4} pc^{-3})$   &  Birth rate $(10^{-13} pc^{-3} yr^{-1})$   &  Caveats   \\ \hline
A        &  2.9 $\pm$ 0.8			&	       5.4 $\pm$ 1.5		     & Not complete	      \\
B        &  3.8 $\pm$ 1.1			&	       5.4 $\pm$ 1.5		     & Too many $E(B-V)$=0	   \\
C        &  3.2 $\pm$ 0.9			&	       7.3 $\pm$ 2.0		     & Too many hot/young, not complete \\
    \hline
\end{tabular} \\ 
}
\end{table*}

\section{Derivation of the space number density of DA white dwarfs from UVEX}
\label{sec:spacedensity}
The observed white dwarf sample from \UVEX is selected from 211 square degrees along the Galactic Plane. 
The simulated numerical sample is obtained from the full Galactic Plane (3\,600 square degrees).
Since the sample with $g$$<$19.5 shows no
longitude or latitude dependence, the area ratio between the observed
sample and the simulated sample is simply a factor 211/3\,600.
Assuming equal depths of 1.0 kpc, the volume of the simulated white dwarf
sample is $0.365$ kpc$^{3}$ and the observed white dwarf sample
is the same factor 3600/211 times smaller.\\

The observed UV-excess samples (A, B, C) contain (276,360,303) sources 
and the simulated sample contains 723 white dwarfs in the full Plane.
If we correct the observed sample for the fraction of genuine DA white dwarfs (67$\%$), there are 
(185,241,203) DA white dwarf candidates in a volume within 1.0 kpc.
If we correct the volume of the observed sample there would be
$723 \times$ (211/3\,600)=42.4 times more theoretical white dwarfs in the volume.
The difference between the number of observed white dwarfs and number
of simulated white dwarfs is a factor (185,241,203)/42.4 = (4.36,5.68,4.79) for cases (A, B, C).\\

Using these ratios to scale the observed sample to the total numerical
sample of $8\times10^4$ white dwarfs with $T_{\rm eff}$$>$10\,000K, 
an average space density in a volume within a radius of 1 kpc around the Sun
is obtained of
$\overline{\rho_A}$ = 2.9 $\pm$ $0.8 \times 10^{-4}$ pc$^{-3}$, 
$\overline{\rho_B}$ = 3.8 $\pm$ $1.1 \times 10^{-4}$ pc$^{-3}$ and
$\overline{\rho_C}$ = 3.2 $\pm$ $0.9 \times 10^{-4}$ pc$^{-3}$.
These results are summarized in Table\ \ref{tab:sdbruvex}.
The derivation of the errors here is explained in Sect.\ \ref{sec:discussion}.\\

To test the validity of the model assumption on Galactic reddening, and
to test the sensitivity of the result as a function of the actual distance/volume used in the calculations, 
the cut-off distance has been varied between 0.1 and 1.0 kpc in steps of $\Delta d$=0.1 kpc. 
The resulting average space
densities are shown in Fig.\ \ref{fig:rhodistance}, which shows that
the result is very stable in the range 0.6 - 1.0 kpc. Below 0.5 kpc
the results change rapidly, both within one sample as well as between
the three samples. This is caused by low number statistics, combined
with the relatively large uncertainties on individual systems,
inherent to the photometric method of deriving temperatures and
reddenings.\\ 

\begin{figure}
\centerline{\epsfig{file=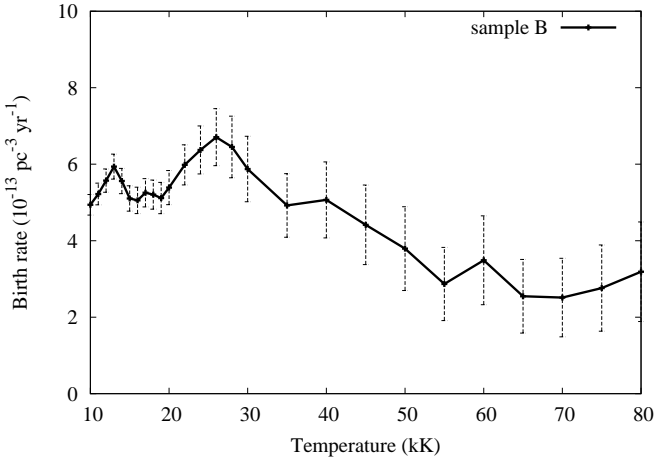,width=9cm,angle=0,clip=}}
\caption{The UVEX birth rate of sample B for different limits of $T_{\rm eff}$.
The error bars, indicating the number of white dwarfs ($N$) used for each birth rate,
are calculated as $1/\sqrt(N)$.
\label{fig:brvstemp}}
\end{figure}

\section{Birth rate of DA white dwarfs in the Galactic Plane}
\label{sec:formationrate}
For the derivation of the birth rate of DA white dwarfs, only objects with $T_{\rm eff}$$\geq$20\,000K are 
taken into account. The samples are limited to $T_{\rm eff}$$\geq$20\,000K
since for the hottest systems the cooling tracks is less uncertain (see Fig.\ \ref{fig:cooling}) and the assumption of a 
constant birth rate is more realistic, while for a higher $T_{\rm eff}$ the number of systems in the samples would become too small.
From the cooling tracks of Fig.\ \ref{fig:cooling} we assume that all white dwarfs in the samples with $T_{\rm eff}$$\geq$20\,000K 
are younger than $\sim 6.9 \times 10^{7}$ years.
From the 211 square degrees of V12a there are (153,154,211) white dwarf candidates with $T_{\rm eff}$$\geq$20\,000K 
for samples A, B and C of Sect.\ \ref{sec:method}, respectively, taking the 67$\%$ into account.
This area of 211 square degrees has a volume of 2.14 $\times$ 10$^{-3}$ kpc$^3$ using a depth of 1.0 kpc.\\ 
 
Within the full numerical sample there are 2\,024 white dwarfs with $T_{\rm eff}$$\geq$20\,000K within 1.0 kpc,
of which 267 have a magnitude $g$$<$19.5 and fall within the simulated sample.
If we correct for the volume and the ratio between simulated and observed 
sources the same way as in Sect.\ \ref{sec:spacedensity}, 
the birth rate for the samples A, B and C is 
(5.4 $\pm$ 1.5) $\times$ 10$^{-13}$ pc$^{-3}$yr$^{-1}$, (5.4 $\pm$ 1.5) $\times$ 10$^{-13}$ pc$^{-3}$yr$^{-1}$ 
and (7.3 $\pm$ 2.0) $\times$ 10$^{-13}$ pc$^{-3}$yr$^{-1}$.
These results are summarized in Table\ \ref{tab:sdbruvex}. 
The derivation of the errors here is explained in Sect.\ \ref{sec:discussion}.\\

So far the birth rate is derived using the samples limited to $T_{\rm eff}$$\geq$20\,000K. 
Fig.\ \ref{fig:brvstemp} shows that birth rate varies between
2.5 $\times$ 10$^{-13}$ pc$^{-3}$yr$^{-1}$ and
6.7 $\times$ 10$^{-13}$ pc$^{-3}$yr$^{-1}$
for different limits of $T_{\rm eff}$.
However, the birth rates at $T_{\rm eff}$$\geq$26\,000K are influenced by the shape of the cooling tracks and 
the age versus absolute magnitude relation (Fig.\ \ref{fig:ageabsolutemag}),
while the birth rates at the higher temperatures are affected by low number statistics.\\

\section{Discussion and conclusions}
\label{sec:discussion}
A derivation of the average space density and birth rate 
within a radius of 1 kpc around the Sun very much depends on the ability to
construct volume-limited samples, to estimate the completeness and
biases in the observational sample, and the accuracy of deriving
fundamental parameters from observational characteristics. As
outlined in Sect.\ \ref{sec:results}, the space densities have
been estimated using three observational samples, each with its own
set of biases and/or corrections. Sample A is a
conservative lower limit, since this sample excludes
white dwarfs left from the unreddened $\log g$=8.0 colour track in the
colour-colour diagram. In sample B all white dwarfs are taken into
account, so the space density derived from this sample is the most
complete, however, the method overestimates the number of systems with no reddening.
For sample C the space density is also a lower limit, since the sample excludes a 
fraction of systems above the grid, while the derived birth rate is too large due to 
an overestimate of the number of hot/young systems.\\

A number of caveats and limitations are in common between the three samples.
The estimated space density depends on: (i) the distance determination, 
(ii) uncertainties in the method for estimating the temperature and reddening
of the white dwarfs in the observational sample, (iii) the assumption about
the amount of reddening/extinction, (iv) the magnitude cut $g$$<$19.5, (v) the
colour cut $(g-r)$$<$0.4, (vi) the assumption of $\log g$=8.0 for all white dwarfs, 
(vii) the fraction of genuine white dwarfs in the UV-excess catalogue and (viii) the binary
fraction in the UV-excess catalogue. The estimated white dwarf
birth rate depends on these points as well, with the additional
assumptions about the cooling time and constant birth rate.\\

The distance estimates to individual systems strongly depend on
the assumed absolute $g$-band magnitudes from Holberg \&
Bergeron (2006), assuming $\log g$=8.0 for all white
dwarfs. The absolute magnitudes follow from the temperature
and reddening determined from the \UVEX photometry. If the
absolute magnitude would be brighter than derived, the white
dwarfs would be detectable over a larger volume. In the most
extreme cases, for example due to a maximal shift in $(U-g)$ of
--0.2 magnitudes, the temperature determinations are off by $\Delta
T_{\rm eff}$=+6\,000K for cool white dwarfs and up to $\Delta T_{\rm
eff}$=+30\,000K for the hottest white dwarfs. The surface gravity
determination could be off by $\Delta \log g$=0.5 (Fig.\,5 of V12b). We
note that although in the last case we would strongly overestimate
an absolute bolometric magnitude, the effect is ameliorated by the
fact that the change in the absolute $g$-band magnitude is less
severe at these very hot temperatures. In these cases the
absolute magnitude would be overestimated by $M_{g}$$\sim$1
magnitude on an individual basis. If the apparent magnitude of
a source at 1.0 kpc would change by $m_{g}$=0.1 magnitude, the distance would typically
change by 5$\%$. If this would be the case for the total sample, this
would mean a maximal increase or decrease of 15$\%$ of the total
survey volume. There might be a Malmquist-type bias in the distance-selected 
observational sample. There will be distance uncertainties
since the observational sample will include white dwarfs which are outside the chosen distance limit, but brought in because of  
distance errors, and it will exclude objects which are moved to outside of the distance limit because of distance errors.
There is no direct effect since the observational sample is compared to the simulated sample, and 
the space densities, calculated using different volumes in Fig.\ \ref{fig:rhodistance}, depend only slightly on the volume. 
For the derivation of an error on the space number density (see below), the effect of this bias is taken into 
account within the factor of 15 per cent of the photometric scatter.\\

The colour cut $(g-r)$$<$0.4 and the magnitude cut $g$$<$19.5
will cause a loss of systems on
the total number of white dwarfs in the observational
sample. In the simulated sample there are three white dwarfs with
$(g-r)$$>$0.4 and $g$$<$19.5: a fraction of $\sim$0.4$\%$,
negligible compared with the other correction factors applied (see Fig.\ \ref{fig:wdhistogramgband}). 
In the observed UV-excess sample there are no sources with $(g-r)$$>$0.4
spectroscopically confirmed as white dwarfs in V12b. 
The probability that a source with $g$$<$19.5 and
$(g-r)$$<$0.4 will be picked-up by the selection algorithm (Fig.\,14
of V12a) drops for sources brighter than $g$=16 and redder than $(g-r)$$>$0.2 to $\sim$50$\%$.
Unreddened white dwarfs cooler than $T_{\rm eff}$$<$7\,000K have synthetic colours
$(g-r)$$>$0.4, due to the colour cut the final white dwarf sample is
incomplete for these cool white dwarfs. However, in the comparison
with the numerical model these cool dwarfs have also been excluded,
and their exclusion from the observational sample therefore does not
influence the estimate on the space density of hotter white dwarfs
(T$>$10\,000 K).\\

The magnitude cut at $g$$<$19.5 is applied due to the difference between the 
magnitude distributions of the simulated sample and the observational samples
for magnitudes fainter than $g$$>$19.5. For fainter magnitudes the number of 
white dwarfs in the simulated sample increases strongly while the number of 
white dwarfs in the observational sample starts to drop. For magnitudes $g$$>$19.5 
the observational sample is not complete which would influence the
result of the space density. If a magnitude limit of $g$$<$20.0 
was chosen, the space densities would have been
(2.4 $\pm$ 0.7) $\times$ 10$^{-4}$ pc$^{-3}$, 
(3.2 $\pm$ 0.9) $\times$ 10$^{-4}$ pc$^{-3}$ and 
(2.7 $\pm$ 0.8) $\times$ 10$^{-4}$ pc$^{-3}$
for the three samples (A, B, C), which is $\sim$16$\%$ smaller.\\

The UV-excess catalogue was selected from 726
partially contiguous`direct' fields, as defined in Gonz\'{a}lez-Solares et al. (2008). 
Because of the tiling pattern of the \IPHAS and \UVEX surveys a completely contiguous area of this number of
fields would result in an overlap in area of $<$5$\%$, which is the maximal
correction on the 211 square degree area that could be applied.
Over the covered area a number of sources might be missed because they fell on dead
pixels or very near the edges of the CCDs. However, the WFC consists
of high quality CCDs and the total dead area is $<$1$\%$.\\

\begin{figure*}
\centerline{\epsfig{file=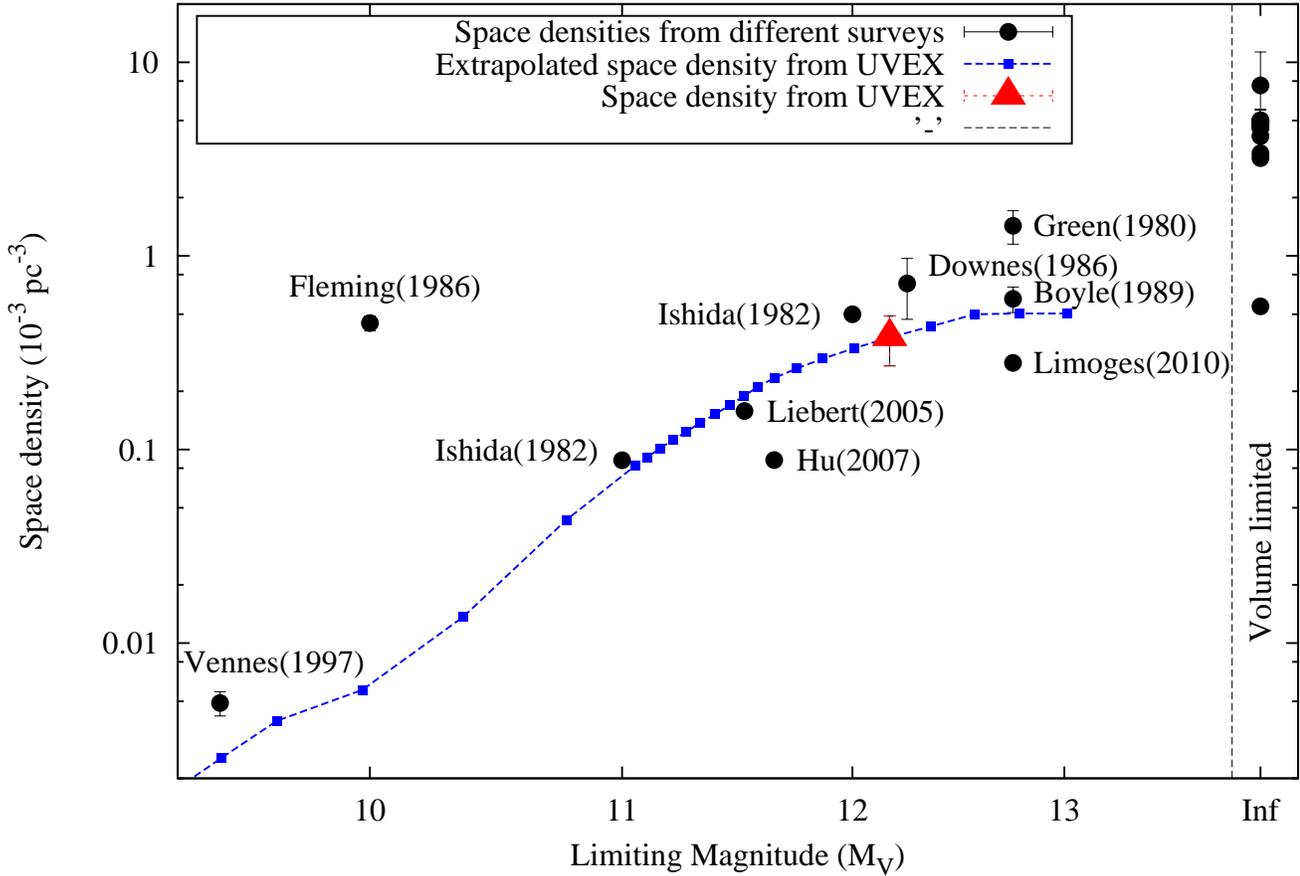,width=18cm,angle=0,clip=}}
\caption{The UVEX space density of sample B (blue squares) extrapolated for different $M_{V}$ overplotted 
on the space number densities from other surveys (black dots). The dots at ``Inf'' represent the space densities
of the volume limited surveys indicated in Table\ \ref{tab:spacedensities}.
\label{fig:sdextrapolated}}
\end{figure*}

An assumption that may strongly affect the estimates is the
assumption of $\log g$=8.0 for all sources. This has been motivated
by the findings in V12b (Fig.\,5) and the well known strong biases
in previous studies of the white dwarf population towards $\log
g$=8.0 (e.g. Fig.\,5 of Vennes et al., 1997 and Fig.\,9 of
Eisenstein et al., 2006.). At face value Fig.\ \ref{fig:ccdcompare}
suggests that in the Plane a substantial number of sources exist
with $\log g <8$, although this is not substantiated by the
spectroscopic fitting in V12b. However, if a large number of lower
gravity systems are present, this would lead to an overestimate of
the space density since lower gravity systems are more luminous at a
given temperature and the observed sample therefore occupies a
larger volume. At a given temperature $\log g$=7.5 gravity 
white dwarfs will have larger absolute magnitudes of $\sim$ 0.8-0.9 mag compared to 
$\log g$=8.0 of DA white dwarfs.
Their distance would be underestimated, and so also the space
density would be overestimated by a factor of $\sim$25$\%$.
For the derivation of an error on the space number density (see below), 
an error on the surface gravity of $\Delta \log g$=0.1, which is a typical value
of the scatter in the white dwarf surface gravity (Fig.\,5 of V12b), will be taken into account.
\\

Combining the uncertainties mentioned above leads to an
upper and lower limit on the space number density derived in
Sect.\ \ref{sec:spacedensity}. If we consider the most optimistic case, the upper limit is due to a combination 
of the method of Sect.\ \ref{sec:method} and photometric scatter of 0.5 mag (15$\%$), 
an error of $\Delta \log g$=0.1 (6$\%$)
non-selected DA white dwarfs at $(g-r)$$>$0.4 (1$\%$), non-selected by the algorithm of V12a (1$\%$) (see Sect.\ \ref{sec:nonselection})
and non-selected white dwarfs due to tiling of the fields and errors on the CCD chips (5$\%$).
The error on the birth rate is estimated in a similar way as for the space density.
If the same uncertainties are taken into account, the birth rate would be 28$\%$ larger in the most optimistic case.
Now for sample C, the distributions are less similar to the
modelled theoretical sample, and the method finds too
many hot solutions. For this reason the birth rate for sample C is larger than for sample B.
\\

\begin{table*}
\caption[]{Space number densities from other surveys. \label{tab:spacedensities} }
\centering
{\small
\begin{tabular}{ | l | c | r | }
    \hline  
Reference    &  Space density $(10^{-3} pc^{-3})$   &  Limits  \\ \hline
UVEX &  0.38$\pm$0.11		      &  $M_{V}<12.2$  \\
Giammichele (2012)$^{*}$ &  4.39                     &  local  \\
Limoges (2010) &  0.280                       &  $M_{V}<12.75$, DA WDs in Kiso  \\
Limoges (2010)$^{*}$ &  0.549                       &  All $T_{\rm eff}$ in Kiso DA WD sample  \\
Sion (2009)$^{*}$    &  4.9$\pm$0.5    	      &  local, 20 pc    \\
Holberg (2008)$^{*}$ &  4.8$\pm$0.5                 &  local, 13 pc (122 WDs)   \\
Holberg (2008)$^{*}$ &  5.0$\pm$0.7                 &  HOS sample   \\
Hu (2007)      &  0.0881                      &  531 SDSS DA WDs, $12kK<T_{\rm eff}<48kK$ ($M_{V}<11.65$) \\
Hu (2007)      &  1.94                        &  531 SDSS DA WDs, $T_{\rm eff}<48kK$		  \\
Harris (2006)$^{*}$  &  4.6$\pm$0.5                 &  local  \\
Liebert (2005) &  0.158                       &  $T_{\rm eff}>13kK$ ($M_{V}<11.52$) \\
Holberg (2002)$^{*}$ &  5.0$\pm$0.7                 &  local, 13 pc  \\
Knox (1999)$^{*}$    &  4.16                        &  local, PM survey	       \\
Tat (1999)$^{*}$     &  4.8                         &  local, 15 pc	      \\
Leggett (1998)$^{*}$ &  3.39                        &  local, $1/V_{max}$  \\
Vennes (1997)  &  0.019$\pm$0.003             &  EUVE sample of 110 DA WDs  \\
Vennes (1997)  &  0.0049$\pm$0.0007           &  hot DA WDs $T_{\rm eff}>40kK$ ($M_{V}<9.45$) \\
Oswalt (1996)$^{*}$  &  7.6$\pm$3.7                 &  local, wide binaries  \\
Weidemann (1991)$^{*}$ & 5                           &  local, in 10pc  \\
Boyle (1989)   &  0.60$\pm$0.09               &  $M_{V}<12.75$   \\
Liebert (1988)$^{*}$ &  3.2                         &  local, $1/V_{max}$  \\
Downes (1986)  &  0.72$\pm$0.25               &  $M_{V}<12.25$   \\
Fleming (1986) &  0.45$\pm$0.04               &  $M_{V}<10$, PG WD sample: 353 obj.  \\
Shipman (1983)$^{*}$ &  4.6                         &  local, astrom. binaries  \\
Ishida (1982)  &  0.088                       &  $M_{V}<11.0$, 588 KUV obj.	\\
Ishida (1982)  &  0.500        	              &  $M_{V}<12.0$, 588 KUV obj.   \\
Green (1980)   &  1.43$\pm$0.28  	      &  $M_{V}<12.75$   \\
Sion (1977)$^{*}$    &  5                           &  local, 23 WDs in 10 pc \\
    \hline
$^{*}$ Volume limited   
\end{tabular} \\ 
}
\end{table*}

\begin{table*}
\caption[]{Birth rates from other surveys. \label{tab:birthrates} }
\centering
{\small
\begin{tabular}{ | l | c | r | }
    \hline
Reference     &  Birth rate $(10^{-13} pc^{-3} yr^{-1})$   &  Limits  \\ \hline
UVEX  &  5.4$\pm$1.5  & $T_{\rm eff}>20kK$   \\
Frew (2008)     &  8$\pm$3      & PN birthrate      \\
Hu (2007)       &  2.579        & $12kK<T_{\rm eff}<48kK$, 531 SDSS DA WDs 		 \\
Hu (2007)       &  2.794        & $T_{\rm eff}<48kK$, 531 SDSS DA WDs  	    \\
Liebert (2005)  &  6            & PG WD sample: 348 obj.      \\
Liebert (2005)  &  10$\pm$2.5   & overall, in local disk	 \\
Holberg (2002)  &  6            & over 8 Gyr  \\
Phillips (2002) &  21           & local PN birth rate  \\
Vennes (1997)   &  8.5$\pm$1.5  & local \\
Pottach (1996)  &  4-80         & local PN birth rate \\
Weidemann (1991) &  23           & derived from star/WD formation model \\
Boyle (1989)    &  $\sim$6      & derived WD birth rate	  \\
Boyle (1989)    &  $\sim$20     & obs. PN birthrate	  \\
Ishida (1987)   &  80           & local PN birth rate \\
Green (1980)    &  20$\pm$10    & from $M_{V}<12.75$ sample          \\
Koester (1977)  &  20           & from $M_{bol}<15.5$ sample     \\
    \hline
\end{tabular} \\ 
}
\end{table*}

Fundamentally the analysis discussed here tests how well the
numerical Galactic model resembles the observed distribution of
white dwarfs. The Galactic model includes an idealized dust
distribution that may not resemble the actual distribution. Since
\UVEX observes directly in the Galactic Plane in blue colours, the
effect of the dust distribution and the ensuing reddening may be
substantial. The theoretical dust distribution in the Sandage
model may behave different than the actual distribution in our
pointings, also because we are looking at a local
population, while the extinction on exactly this local scale is
very poorly known (Sale et al., 2009 and Giammanco et al.,
2011). As can be seen in Fig.\ \ref{fig:wdhistogramav} there
is a difference between the reddening distributions, 
which is partly due to the crude determination of
$E(B-V)$ for the observational sample, with bins of $\Delta
E(B-V)$=0.1. The effect of reddening for the white dwarfs in \UVEX
was already shown in Fig.\,8 of V12b. 
The reddening is smaller than
$E(B-V)$=0.7 ($A_{V}$$<$2.2) for all white dwarfs as shown in Fig.\ \ref{fig:wdhistogramav}
and Fig.\,8 of V12b. In the simulated observable sample there are no sources with
$E(B-V)$$>$0.7. The most reddened white dwarfs have
$E(B-V)$=0.55. In the observational sample the reconstruction method
of Sect.\ \ref{sec:method} finds $E(B-V)$=1.0 and $T_{\rm eff}$=14kK
for only one source, five sources with $T_{\rm eff}$$>$40kK have
$E(B-V)$=0.7 and all other sources have $E(B-V)$$<$0.7.\\

When the local population of white dwarfs is
well-known and spectroscopically characterized it can conversely
be used to derive a 3D extinction map of the local ($d<1$kpc)
environment.\\

Finally, we note that no correction has been made for the binary
fraction of systems dominated by a DA white dwarf in the UV-excess catalogue. 
The binary fraction estimates range from 12$\%$ to 50$\%$
(e.g. Nelemans et al., 2001, Han 1995, Miszalski et al., 2009 and
Brown et al., 2011). The space density and birth rate number derived
here are therefore DA white dwarf dominated systems that fall within
our colour selection criteria, including an
unknown binary fraction.\\

\subsection{Comparison with other surveys}
\label{sec:comparison}
The space density of (3.8 $\pm$ 1.1) $\times$ 10$^{-4}$ pc$^{-3}$, derived for sample B for 
white dwarfs with $M_{V}$$<$12.2 or $T_{\rm eff}$$>$10\,000K, and a birth rate of 
(5.4 $\pm$ 1.5) $\times$ 10$^{-13}$ pc$^{-3}$yr$^{-1}$ over the last 7$\times$10$^7$ years, 
can be compared with the results of other surveys
(Tables\ \ref{tab:spacedensities} and\ \ref{tab:birthrates}). 
All previous estimates have been obtained either from bright samples (in
particular the early surveys) and/or at high Galactic latitudes. The
current study is the first to be obtained in the Galactic Plane
itself where the majority of systems resides. 
For that reason, and due to different magnitude limits, it is not possible 
to automaticaly compare different surveys. As is evident from 
Tables\ \ref{tab:spacedensities} and\ \ref{tab:birthrates}, 
the estimates on the space densities and birth rates strongly vary.
To compare our results to those of the other studies, the Galactic model 
is used to calculate the effective space density at various limiting magnitudes, calibrated to our
result. Fig.\ \ref{fig:sdextrapolated} shows that, when a correction is made for the vaying limiting 
absolute magnitude, many surveys are quite consistent with each other,
despite the fact that they observe different white dwarf samples.
Surveys that claim to be volume limited derive an average space density of $\sim$4.6 $\times$ 10$^{-4}$ pc$^{-3}$,
which is consistent with the extrapolated, continued slope of the space density as a function of absolute 
magnitude of Fig.\ \ref{fig:sdextrapolated}.
However, it is not possible to extrapolate the \UVEX-based result to the coolest/faintest systems since
the star formation history in the Galaxy will start to play a dominant role.
For the same reason also the birth rate results of the different surveys 
in Table\ \ref{tab:birthrates} strongly vary.
If we compare the birth rate of 5.4 $\pm$ 1.5 $(10^{-13} pc^{-3} yr^{-1})$, 
derived for sample B in this paper, the result of \UVEX is consistent with other
estimates.\\

\section*{Acknowledgements}
This paper makes use of data collected at the Isaac Newton Telescope,
operated on the island of La Palma by the Isaac Newton Group in the
Spanish Observatorio del Roque de los Muchachos of the Inst\'{\i}tuto
de Astrof\'{\i}sica de Canarias. The observations were processed by
the Cambridge Astronomy Survey Unit (CASU) at the Institute of
Astronomy, University of Cambridge. Hectospec observations shown in
this paper were obtained at the MMT Observatory, a joint facility of
the University of Arizona and the Smithsonian Institution.
KV is supported by a NWO-EW grant
614.000.601 to PJG and by NOVA.  The authors would like to thank
Detlev Koester for making available his white dwarf model spectra.  
The colour tables and model calculations of
Pierre Bergeron can be obtained from this website
(http://www.astro.umontreal.ca/$\sim$bergeron/CoolingModels) and are
explained in Holberg $\&$ Bergeron (2006, AJ, 132, 1221), Kowalski
$\&$ Saumon (2006, ApJ, 651, L137), Tremblay et al. (2011, ApJ, 730,
128), and Bergeron et al. (2011, ApJ, 737, 28).\\

\label{lastpage}

\newpage

\appendix

\section{Recalibrated UVEX data.}
\label{app:recalibrated}
\begin{figure}
\centerline{\epsfig{file=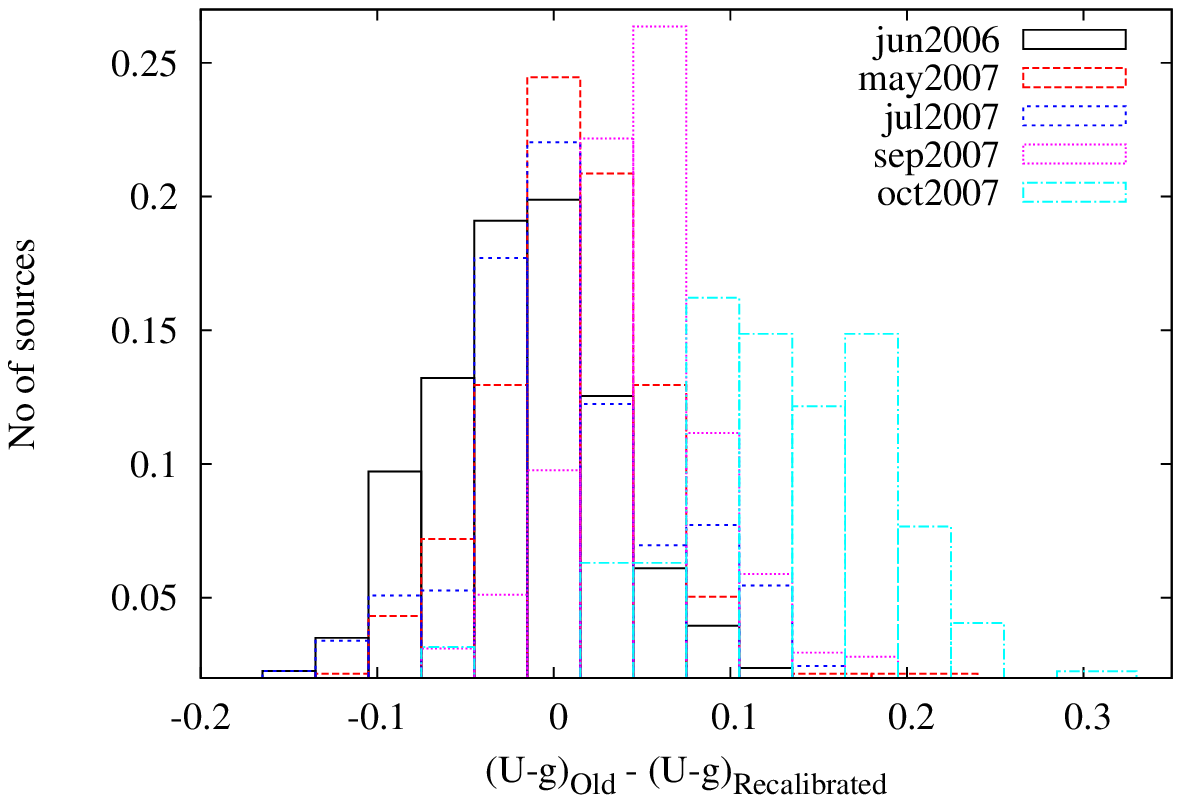,width=9cm,angle=0,clip=}}
\caption{Difference in $(U-g)$ between the original UVEX data and recalibrated UVEX data for the 5 different months used for 
the UV-excess catalogue of V12a.
\label{fig:ugpermonth}}
\end{figure}

There is a possible systematic shift in the original UV-excess catalogue $(U-g)$ data, which would influence 
the result of methods in Sect.\ \ref{sec:method}. For that reason we use recalibrated \UVEX data, as explained in Greiss et al. (2012). 
The differences in $(U-g)$ between the original \UVEX data and recalibrated \UVEX data for the 5 different months used in 
V12a are plotted in Fig.\ \ref{fig:ugpermonth}.
The shift in the original \UVEX data does not influence the content of the UV-excess 
catalogue because the selection in V12a was done relative to the reddened main-sequence population.
The magnitudes and colours of the UV-excess sources might still show a small
scatter, similar to the early \IPHAS data (Drew et al., 2005), since a global photometric calibration is not applied to the \UVEX data yet.\\

\section{Distributions of simulated and observational samples}
\label{app:histograms}

\begin{figure*}
\centerline{\epsfig{file=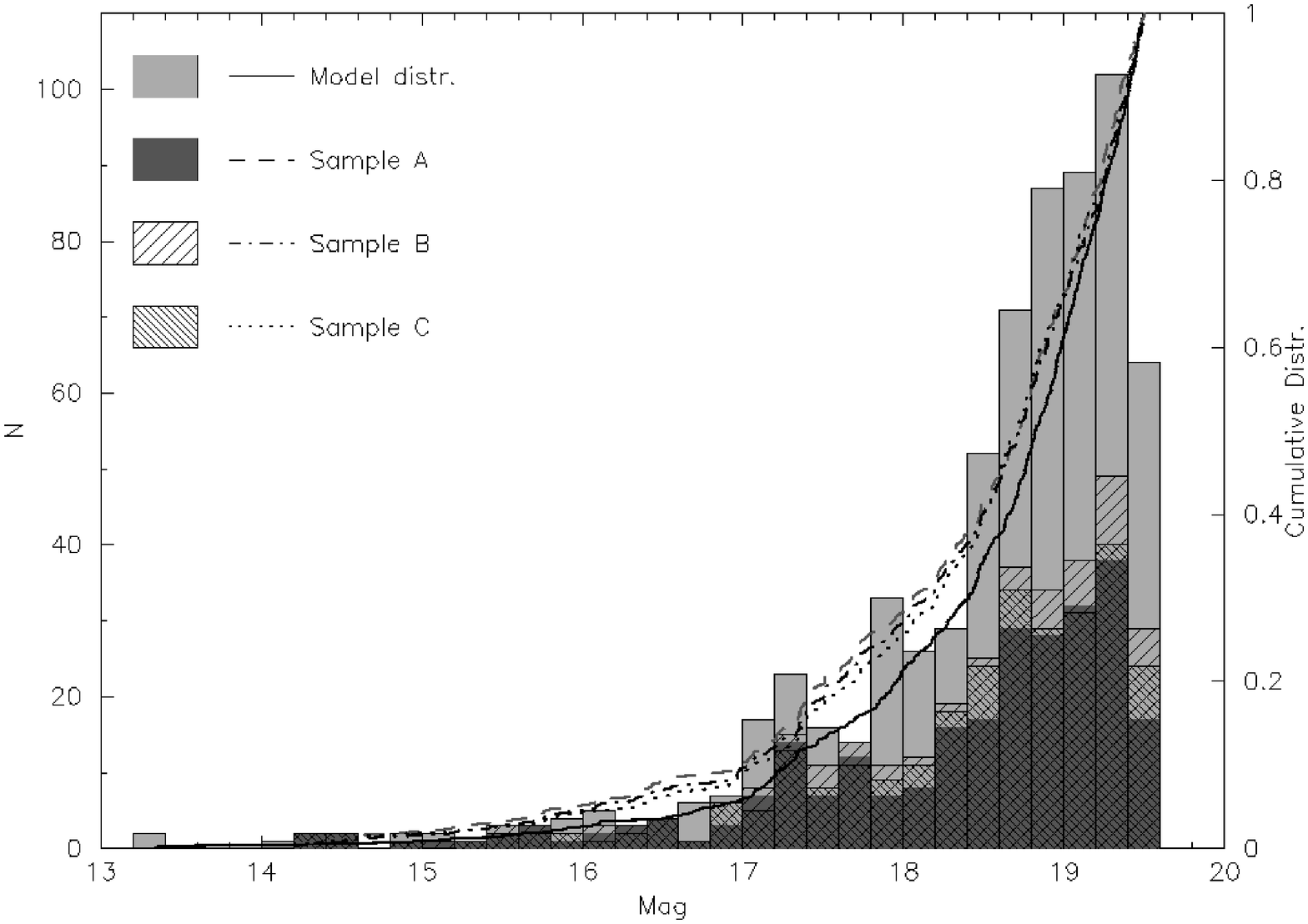,width=10cm,angle=-90,clip=}}
\caption{Magnitude distributions and cumulative distributions of the UV-excess white dwarf
  candidates from the 3 samples A-C and the simulated white dwarf sample.
\label{fig:magn3samples}}
\end{figure*} 

\newpage

\begin{figure*}
\centerline{\epsfig{file=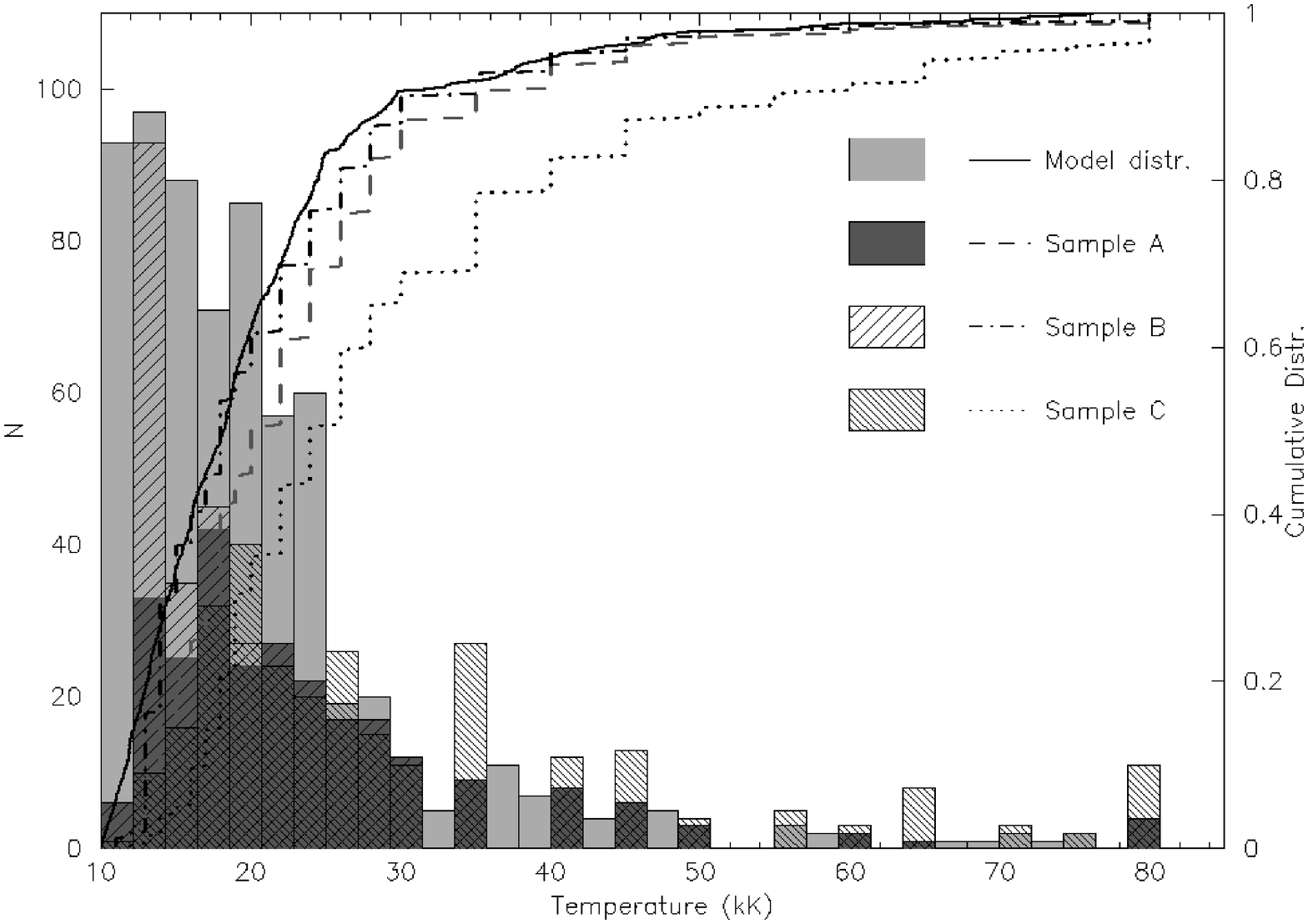,width=10cm,angle=-90,clip=}}
\caption{Temperature distributions and cumulative distributions of the UV-excess white dwarf
  candidates from the 3 samples A-C and the simulated white dwarf sample.
\label{fig:temp3samples}}
\end{figure*}

\newpage

\begin{figure*}
\centerline{\epsfig{file=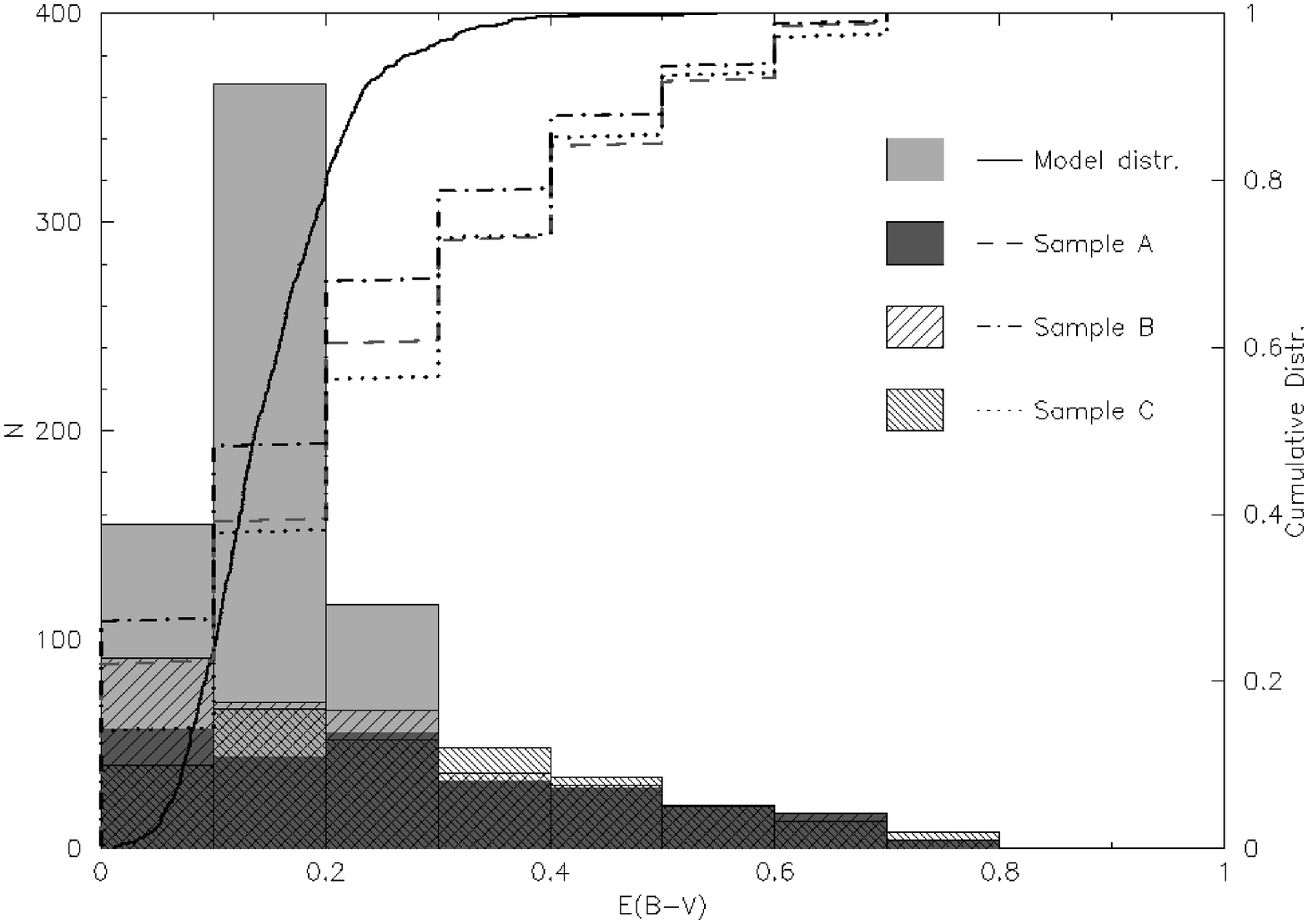,width=10cm,angle=-90,clip=}}
\caption{Reddening distributions and cumulative distributions of the UV-excess white dwarf
  candidates from the 3 samples A-C and the simulated white dwarf sample.
\label{fig:redd3samples}}
\end{figure*}

\newpage

\begin{figure*}
\centerline{\epsfig{file=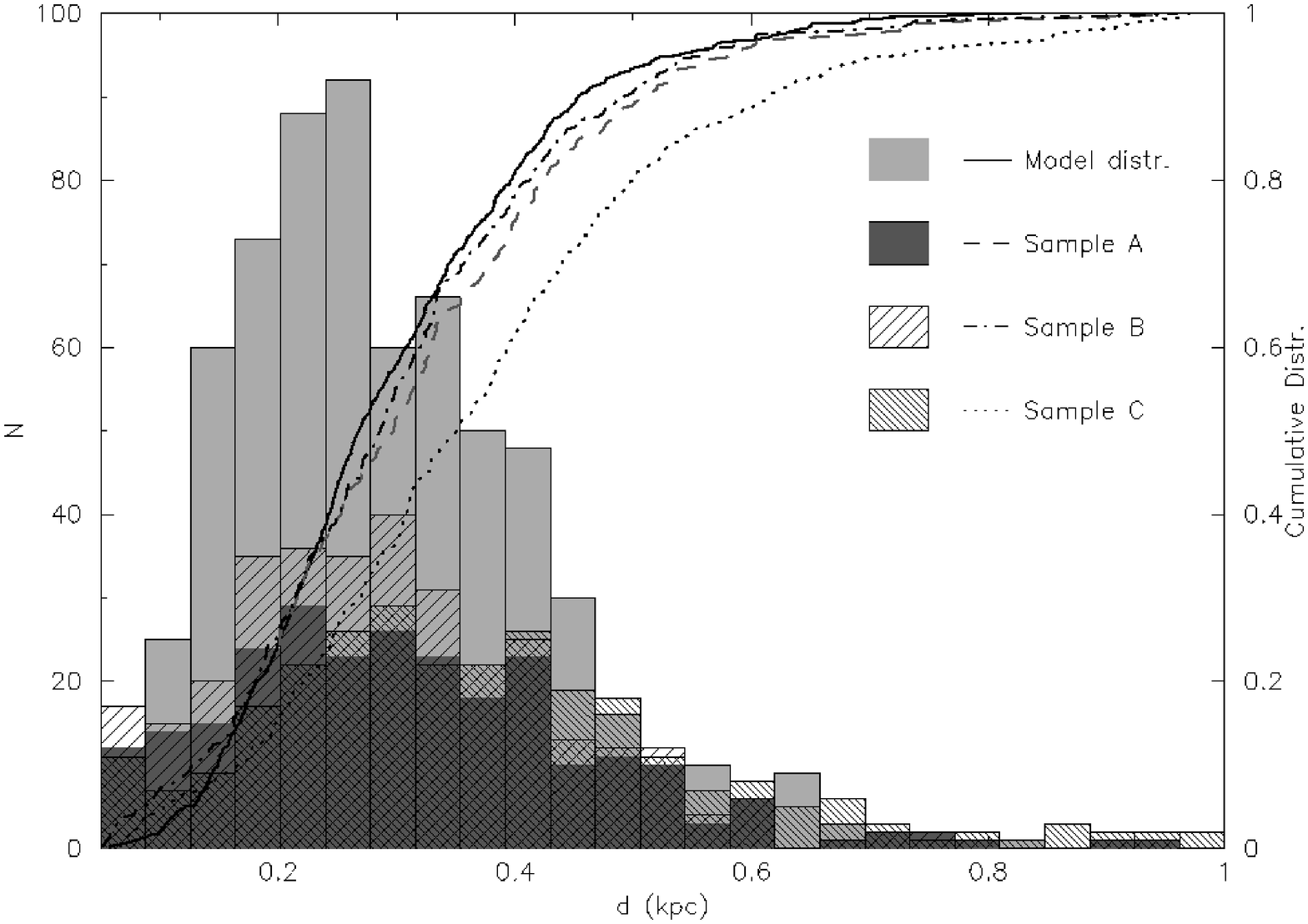,width=10cm,angle=-90,clip=}}
\caption{Distance distributions and cumulative distributions of the UV-excess white dwarf
  candidates from the 3 samples A-C and the simulated white dwarf sample.
\label{fig:dist3samples}}
\end{figure*}

\label{lastpage}

\end{document}